\documentclass[11pt,reqno]{article}
\usepackage{amsmath}
\usepackage{amsfonts}
\usepackage{amssymb}
\usepackage{hyperref}
\usepackage{latexsym}
\usepackage[dvips]{graphicx}
\usepackage{epsf}
\usepackage{color}

\allowdisplaybreaks \numberwithin{equation}{section}

\newcommand{\be}{\begin{equation}}
\newcommand{\ee}{\end{equation}}
\newcommand{\bea}{\begin{eqnarray}}
\newcommand{\eea}{\end{eqnarray}}

\newcommand{\f}{\frac}

\let\a=\alpha \let\b=\beta    
\let\z=\zeta     \let\th=\theta   \let\l=\lambda
\let\m=\mu    \let\n=\nu          \let\r=\rho 
\let\s=\sigma      
    
\let\G=\Gamma \let\D=\Delta   \let\L=\Lambda 
         
  \let\eps=\epsilon

\let\==\equiv

\newcommand{\p}{\partial}
\newcommand{\na}{\nabla}
\newcommand{\Tr}{{\rm Tr}}

\newcommand{\gb}{\bar{g}}
\newcommand{\hb}{\bar{h}}

\newcommand{\Gb}{\bar{\Gamma}}
\newcommand{\cb}{\bar{c}}
\newcommand{\Cb}{\bar{C}}
\newcommand{\phib}{\bar{\phi}}

\newcommand{\ft}{\tilde{f}}
\newcommand{\tG}{\tilde{G}}
\newcommand{\Lt}{\tilde{\L}}
\newcommand{\Rt}{\tilde{R}}

\newcommand{\cC}{\mathcal{C}}

\newcommand{\cL}{\mathcal{L}}

\newcommand{\cN}{\mathcal{N}}

\newcommand{\cP}{\mathcal{P}}
\newcommand{\cR}{\mathcal{R}}
\newcommand{\cS}{\mathcal{S}}
\newcommand{\cT}{\mathcal{T}}

\newcommand{\ug}{\underline{g}}

\begin{document}

\thispagestyle{empty}
\begin{flushright} \small
AEI-2012-017\\
PI-QG-260
\end{flushright}
\bigskip

\begin{center}
 {\LARGE\bfseries   The local potential approximation in quantum gravity}\\
[10mm]
Dario Benedetti$\ ^{1,*}$ and Francesco Caravelli$\ ^{1,2,3,\dagger}$
\\[3mm]
{\small\slshape
$\ ^1$Max Planck Institute for Gravitational Physics (Albert Einstein Institute), \\
Am M\"{u}hlenberg 1, D-14476 Golm, Germany \\ \vspace{.3cm} 
$\ ^2$Perimeter Institute for Theoretical Physics, \\
Waterloo, Ontario N2L 2Y5
Canada  \\ \vspace{.3cm} 
$\ ^3$ University of Waterloo, Waterloo, Ontario N2L 3G1, Canada \\
\vspace{.3cm}
$^*$ {\upshape\ttfamily dario.benedetti@aei.mpg.de} \\
$^\dagger$ {\upshape\ttfamily fcaravelli@perimeterinstitute.ca}
} 

\end{center}
\vspace{5mm}

\hrule\bigskip

\centerline{\bfseries Abstract} \medskip
Within the context of the functional renormalization group flow of gravity, we suggest that a generic $f(R)$ ansatz 
(i.e. not truncated to any specific form, polynomial or not) for the effective action plays a role analogous to the local potential approximation (LPA) in scalar field theory. 
In the same spirit of the LPA, we derive and study an ordinary differential equation for $f(R)$ to be satisfied by a fixed point of the renormalization group flow.
As a first step in trying to assess the existence of global solutions (i.e. true fixed point) for such equation, we investigate here the properties of its solutions by a comparison of various series expansions and numerical integrations.
In particular, we study the analyticity conditions required because of the presence of fixed singularities in the equation, and we develop an expansion of the solutions for large $R$ up to order $N=29$. Studying the convergence of the fixed points of the truncated solutions with respect to $N$, we find a characteristic pattern for the location of the fixed points in the complex plane, with one point stemming out for its stability. 
Finally, we establish that if a non-Gaussian fixed point exists within the full $f(R)$ approximation, it corresponds to an $R^2$ theory.

\bigskip
\hrule\bigskip
\newpage
\tableofcontents

\section{Introduction}

The application of functional renormalization group techniques to gravity has generated many appealing results in support of the asymptotic safety scenario \cite{Weinberg:1980gg,Niedermaier:2006wt,Percacci:2007sz,Litim:2008tt,Reuter:2012id}.
The main tool used in such investigations is the so-called Functional Renormalization Group Equation (FRGE) for the effective average action $\G_k$ \cite{Wetterich:1992yh}, which reads\footnote{Here $\Phi$ denotes the collection of all the fields in the theory under consideration, and ${\rm STr}$ a functional supertrace over their spinorial indices and spacetime coordinates (collectively denoted by the $A,B$ indices). The running scale is $t=\ln k$, and $\cR_k$ is a cutoff function implementing the Wilsonian momentum-shell integration. For further details we refer to the many general reviews \cite{Morris:1998da,Bagnuls:2000ae,Berges:2000ew,Pawlowski:2005xe,Gies:2006wv,Delamotte-review}.}
\be\label{FRGE}
\f{d}{dt} \G_k[\Phi] = \f12 {\rm STr} \left[ \left( \frac{\delta^{2} \G_k}{\delta \Phi^A \delta \Phi^B } + \cR_k \right)^{-1} \, \f{d}{dt}  \cR_k  \right] \, .
\ee
The FRGE is an exact equation whose solutions determine a flow of effective actions in the theory space of all possible functionals  $\G_k$, interpolating between a bare action at some initial UV scale $k=\L$ and the full effective action at $k=0$.
The primary goal of the asymptotic safety program is to show that there exists a non-trivial (i.e. non-free) gravitational action $\G^*$ which is a fixed point\footnote{That is, $\p_t \G^*=0$. Here $\p_t$ is a partial derivative, whereas in \eqref{FRGE} $d/dt$ is a total derivative, the difference being that the latter acts also on the scaling dimension of the fields and of the Lagrangian itself. This point will be made more explicit in \eqref{lhs}.} for such a flow, with finitely many relevant directions.

Due to the complexity of the question, the main line of progress in this topic has been based on a sort of ``mathematical experiments'': an equation, which is in principle exact, is solved by truncating its infinite-dimensional functional domain to a finite-dimensional subspace; such operation is repeated for various truncations, 
and stability and convergence of the results are tested.
In practice, exploring larger and larger truncations is a very tough and tedious job, and the question is how confident can we be about our conclusions, based on the results we have obtained so far.

Such a question is of course not specific to gravity, but common to any application of exact renormalization group equations in which the approximation scheme does not rely on a small parameter expansion, but rather on an apparently arbitrary truncation of the space of action functionals.
In simpler settings than gravity, as for scalar field theory, the use of the FRGE has reached a sufficient level of confidence that allows us to make very solid statements about the phase structure of the theory and even quantitative predictions that can compete with other methods (see for example \cite{Litim:2010tt} and references therein).
One essential element for such achievement is the use of
the RG equations as differential equations for unspecified functionals, rather than for the couplings of a usual field expansion. 
For instance, rather than truncating the scalar potential $V_k(\phi)$ to a polynomial of order $N$, looking at the RG flow of the coefficients of the various monomials and studying what happens as $N$ is increased, one can instead view the RG flow equation as a partial differential equation for the unknown function $V_k(\phi)$. The lowest order of approximation in such a scheme is called Local Potential Approximation (LPA) and consists in retaining only the equation for $V_k(\phi)$. At next order one includes the equation for a ``wave function renormalization functional'' $Z_k(\phi)$, and so on.
Explicitly, one writes a derivative expansion of the type
\be \label{deriv-exp}
\begin{split}
\G_k[\phi] = & \int d^d x \Big[ V_k(\phi) + Z_k(\phi) \p^\m \phi \p_\m\phi   \\
& + W^a_k(\phi) (\p^2\phi)^2 + W^b_k(\phi) \p^\m \phi \p_\m\phi (\phi \p^2\phi) + W^c_k(\phi) (\p^\m \phi \p_\m\phi)^2 +O(\p^6)  \Big] \, ,
\end{split}
\ee
which plugged into \eqref{FRGE} leads to partial differential equations for the unknown functions $V_k(\phi)$, $Z_k(\phi)$, etc.
A comparative study between such scheme and the polynomial truncations was carried out by Morris in \cite{Morris:1994ki}, highlighting the greater reliability of the former,
which leads to more accurate results, and to a better understanding of the convergence properties of truncations.
Of course polynomial truncations are still used because they are easier to handle, and they can provide accurate results in many cases (e.g. \cite{Canet:2003qd}), however the validity of such truncations is better assessed by a comparison to the derivative expansion.

In the case of pure gravity most of the truncations studied so far take the form\footnote{Details of the contruction of truncations in the gravitational case will be reviewed in Sec.~\ref{Sec:general}.}
\be \label{poly}
\G_k[g_{\m\n}] = \int d^d x\sqrt{g} \sum_{i=0}^N u_i(k) R^i \, ,
\ee
where $R$ is the Ricci scalar, and the truncation order has been increased from the original $N=1$ \cite{Reuter:1996cp,Dou:1997fg,Lauscher:2001ya,Litim:2003vp}, and $N=2$ \cite{Granda:1998wn,Lauscher:2002mb}, to $N=6$ \cite{Codello:2007bd,Machado:2007ea}, $N=8$ \cite{Codello:2008vh} and $N=10$ \cite{Bonanno:2010bt}.
In \cite{BMS1} an $R_{\m\n\r\s}R^{\m\n\r\s}$ term was added also to the $N=2$ truncation, and in \cite{Machado:2007ea} some non-analytic terms like $R^{-1}$ and $\ln R$ have been included.\footnote{Matter has also been included \cite{Percacci:2003jz,BMS2}, as well as corrections to the ghost sector \cite{Groh:2010ta,Eichhorn:2010tb} and bimetric truncations \cite{Manrique:2009uh,Manrique:2010am}. Here we will concentrate on single-metric pure gravity sector of the theory.}

Organizing the truncations of the gravitational action in powers of curvature is a natural starting point, and one familiar in effective field theory \cite{Donoghue:1994dn}.
Such truncations do not correspond of course to polynomial truncations in the fundamental field, which is the metric, and which appears non-polynomially in the curvature, thus differing in this respect from the polynomial truncations in the scalar field theory case.
On the other hand, just as in that case, polynomial truncations lead to ordinary differential equations for the flow of a finite number of couplings, and to algebraic equations for the fixed points. Furthermore, not all higher-order invariants in the curvature contribute with higher-order derivatives to the two-point function, thus the expansion in powers of the curvature is in this sense also not a derivative expansion.

In the present work we advocate the point of view that truncations of the type \eqref{poly} are akin to the polynomial truncations in scalar field theory,
and we propose an analogy between a generic $f(R)$ approximation and the LPA.
Of course, due to diffeomorphism invariance, no potential can be written just for the metric fluctuations, and instead the simplest Lagrangian that can be written without restricting to any specific function is that of an $f(R)$ theory. That is, we retain a generic dependence of the action on the scalar curvature $R$, but discard any derivatives of $R$ as well as any more complicated tensorial structures like $R_{\m\n\r\s}R^{\m\n\r\s}$, etc.
It is well known that such a theory has only one additional (scalar) degree of freedom with respect to general relativity \cite{Sotiriou:2008rp,DeFelice:2010aj}, or in other words it only adds fourth-order derivatives to the trace component of the metric fluctuations, which in general relativity is not a dynamical degree of freedom.
Fourth-order derivatives for the transverse-traceless sector of the metric would be contained for example in terms like $f_2( R_{\m\n\r\s}R^{\m\n\r\s})$, while higher derivatives for both sectors would start for example with terms like $Z_1(R) R \na^2 R$ or  $Z_2(R) R^{\m\n} \na^2 R_{\m\n}$. 
We will not attempt here to define a full derivative expansion to higher orders, however it is clear that the $f(R)$ functional is the functional ansatz with the minimal number of derivatives (of course among those containing an arbitray function, Einstein-Hilbert or Gauss-Bonnet being special cases),
and hence it can serve as the leading term in such an approximation scheme.

There is also another analogy that can be drawn between the LPA and the $f(R)$ ansatz.
In the evaluation of \eqref{FRGE} one first expands an action functional like \eqref{deriv-exp} around a background field $\phib$, writing $\phi=\phib+\varphi$ and computing the second variation with respect to $\varphi$, then one plugs the result into \eqref{FRGE} and projects it on the background, setting $\varphi=0$.
In light of that, we can restate the LPA as being the approximation in which we take a constant background $\phib$: once we project the FRGE on such a background, clearly the only functional we can discern is a potential.
Analogously, in the gravitational case once we choose a maximally symmetric background (typically of spherical topology), as it is usually done in order to keep the calculations manageable, the only functional we can discern is an $f(R)$.
Indeed for a spherical background  the Weyl tensor is identically zero, and the Riemann and Ricci tensors are both proportional to the Ricci scalar $R$, which is constant.
Note that given the approximation consisting in the choice of a spherical background, no further approximations are needed.
In particular no truncation is needed, as we can write down a generic action which contains all possible terms that might ever be generated by the functional traces
of the FRGE on a spherical background: this is the $f(R)$ action.

In the present work we will study the gravitational FRGE in the $f(R)$ approximation, in the spirit of the LPA. We will derive the ordinary differential equation to be satisfied by a fixed-point $f(R)$ function, and we will study general properties of its solutions. As pointed out in \cite{Hasenfratz:1985dm} for the LPA and emphasized in \cite{Morris:1994ki}, most of the solutions to the fixed-point equation end at a singularity, while a putative fixed point should be represented by a non-singular solution.\footnote{Besides being a reasonable physical requirement, it was also proved by Felder \cite{Felder:1987}, within the LPA, that any fixed point which is the limit for $t\to\infty$ of an effective potential with non-singular initial condition at $t=0$ must be a global solution of the fixed-point equation.}
For our equation the identification of global solutions turns out to be a very challenging task, and we report here the present status of our understanding, postponing to future work a more comprehensive numerical study of the solutions. 
Here we will examine in some detail the different type of singularities that plague our equation, identify two analyticity conditions to be satisfied by the solutions, and develop a new series expansion, for large $R$, which appears to be much more manageable than the usual expansion at $R=0$.
The main practical outcomes of our work are an identification of candidate fixed points from the expansion of the solutions at large $R$, and the observation that if a non-trivial fixed point exists, it necessarily corresponds to an $R^2$ theory.

In Sec.~\ref{Sec:general} we present more precisely the $f(R)$ ansatz, and discuss the main results of the paper. The reader who is not interested in the technical details of the computation can in principle skip directly to the conclusions after reading Sec.~\ref{Sec:general}. All the technical details will be presented in the remaining sections: in Sec.~\ref{Sec:Hessian} we provide the functional variations of the action functional, together with the ghost and auxiliary sectors; in Sec.~\ref{Sec:cutoff} we present the cutoff scheme employed in our computation; in Sec.~\ref{Sec:sums} we describe the method used in evaluating the functional traces. Finally in Sec.~\ref{Sec:FPeq} we collect the results into the final form of our FRGE, and in Sec.~\ref{Sec:analysis} we detail the analysis of the fixed-point equation.
A summary of results and future prospects ends the paper in Sec.~\ref{Sec:concl}.

\section{The $f(R)$ approximation: setup and outline of results}
\label{Sec:general}

The flow equation \eqref{FRGE} is adapted to the gravitational case along the lines of the standard field-theoretic quantization, as used also in one- and two-loop calculations \cite{'tHooft:1974bx,Goroff:1985th}. Differences only reside in the cutoff choice, and in the approximation being used.

In the case of pure gravity, the fields comprise the metric, the ghosts and occasionally some auxiliary fields implementing the functional Jacobians originated by field redefinitions.
We use the background field method to obtain a gauge-invariant average effective action, and we define the decomposition of the metric by
\be \label{backgr}
\underline{g}_{\m\n} = g_{\m\n}+h_{\m\n} \, ,
\ee
with $g_{\m\n}$ denoting the background and $h_{\m\n}$ the fluctuations. As background metric we will take a $d$-dimensional sphere, as explained in the introduction.\footnote{For what follows, all one needs to know about the sphere is that the Ricci scalar is constant, $R_{\m\n}=\f{1}{d}g_{\m\n}R$, $R_{\m\n\r\s}=\f{1}{d(d-1)}(g_{\m\r}g_{\n\s}-g_{\m\s}g_{\n\r})R $, and that the radius and the volume are given by $\r^2=d(d-1)/R$ and $V=(4\pi \r^2)^{d/2} \G(d/2)/\G(d)$, respectively. We will also need the eigenvalues of the Laplacian, reported in Table~\ref{table}.}

Following  \cite{Reuter:1996cp}, it is useful to cast a general truncation of the effective average action in the following form:
\be \label{fullGamma}
\G_k[\underline{\Phi},\Phi] = \Gb_k[\ug] + \widehat{\G}_k[h, g] + 
\G_{\rm gf}[h, g] + \G_{\rm gh}[h, g, {\rm ghosts}] + S_{\rm aux}[g, {\rm aux.fields}] \, .
\ee
 In this decomposition $\bar{\Gamma}_k[\ug]$ depends only on the total metric, and it is the proper gravitational action. $\G_{\rm gf}$ and $\G_{\rm gh}$ denote the gauge-fixing and ghost-terms respectively, for which we will take the classical functionals but eventually allowing a running of the gravitational couplings, while $S_{\rm aux}$ is a coupling-independent quadratic action encoding the Jacobians. $\widehat{\Gamma}_k[h, g]$ vanishes for $h=0$, and it encodes the deviations from standard Ward identities due to the use of a cutoff \cite{Reuter:1996cp}. The role of such term has been investigated via bimetric truncations in \cite{Manrique:2009uh,Manrique:2010am}. In the present work we will make use of the common approximation $\widehat{\Gamma}_k=0$; in such case, it suffices to study the FRGE at $h_{\m\n}=0$.

Our ansatz for the gravitational action is
\be \label{ansatz1}
\Gb_k = Z_k \int d^d x\sqrt{\ug} f_k(\underline{R}) 
 \, ,
\ee
where $Z_k=(16\pi G_k)^{-1}$ and $G_k$ is the (running) Newton's constant. 
The running RG scale is $k$, and we will introduce the dimensionless Ricci scalar $\Rt \equiv R/k^2$, and Lagrangian
\be
\ft_k (\Rt) = \f{k^{-d}}{16\pi G_k} f_k(k^2\Rt) \, ,
\ee
whose shape we will try to fix by use of the FRGE.

When plugging our ansatz (\ref{fullGamma}-\ref{ansatz1}) into \eqref{FRGE} we will obtain a partial differential equation for $\ft_k( \Rt)$.
We can trivially write down the left-hand side of the FRGE:
\be \label{lhs}
\f{d}{dt}  \G_k {}_{\big{|}_{h_{\m\n}={\rm ghosts}=0}}= k^d  \int d^d x\sqrt{g} \left\{ \p_t \ft_k(\Rt) + d \ft_k(\Rt) - 2 \Rt \ft_k'(\Rt)    \right\} \, .
\ee
Note that $\p_t$ here is a partial derivative acting only on the explicit $k$-dependence in the function $\ft_k(\Rt)$, while primes denote differentiation with respect to $\Rt$.
The right-hand side of the equation is also a function of $\ft(\Rt)$ and its derivative with respect to both $k$ and $\Rt$. Hence the FRGE is a partial differential equation (PDE) for the function $\ft(\Rt)$.

Now remember that a step of the RG flow consists of two intermediate steps \cite{Bagnuls:2000ae}: integration over a momentum shell, which we take care of via the FRGE; and rescaling of all dimensionful quantities to restore the original UV cutoff. As we have translated everything into dimensionless quantities, the second step is trivially taken care of,
and hence a fixed point of the RG flow has to satisfy
\be \label{FP-condition}
\p_t  \ft_k(\Rt)  = 0 \, .
\ee
As a result, the PDE obtained from the FRGE will reduce to an ordinary differential equation (ODE) at the fixed point. In contrast, in polynomial truncations one gets a system of ordinary differential equations for the flow, and a system of algebraic equations for the fixed point. As explained in the intro, the price we have to pay for going to a higher level of equations should be compensated by the gain in confidence in the results thus obtained.

We are now going to highlight the main results of our analysis, and anticipate some conclusions. All the details will be given in the following sections.

The study of the FRGE in the described approximation presents two challenges: the actual evaluation of the r.h.s. of the equation\eqref{FRGE}, and the numerical study of the PDE obtained in this way.
The first part will be described in sections \ref{Sec:Hessian}, \ref{Sec:cutoff}, \ref{Sec:sums} and \ref{Sec:FPeq}.  A similar calculation has been done before \cite{Machado:2007ea,Codello:2008vh}, but we will adopt here a different implementation of the ghost sector, a different cutoff scheme and a different evaluation technique for the traces, mostly following \cite{Benedetti:2011ct}.
The main result of this part of our work is hence is the derivation of the fixed-point equation in $d=4$ dimensions, which takes the form\footnote{We denote the function at the fixed point just by omitting the subscript $k$ from $\ft$.} 
\be \label{FP-eq}
\ft'''(\Rt) = \frac{\cN( \ft,\ft',\ft''; \Rt)}{ \Rt (\Rt^4-54 \Rt^2-54) \left(( \Rt-2 ) \ft'(\Rt) -2 \ft(\Rt)\right)} \, ,
\ee
where
$\cN( \ft,\ft',\ft''; \Rt)$ is a polynomial in $\ft(\Rt)$ and its first two derivatives, with coefficients polynomial in $\Rt$. Its precise expression will be given later in \eqref{eq-numerator}.

For the second part (Sec.~\ref{Sec:analysis}), we will restrict to the question of existence of fixed points, thus reducing the task to the study of the ordinary differential equation \eqref{FP-eq}. 
We will address in particular the following topics:\\
\indent 1. fixed singularities, initial conditions and analyticity conditions;\\
\indent 2. movable singularities;\\
\indent 3. large-$\Rt$ expansion (boundary condition at infinity);\\
\indent 4. fixed points from truncations of the large-$\Rt$ expansion;\\
\indent 5. generic form of a fixed-point effective action.

As the solution of a third-order ODE is specified by three initial conditions, it would seem at first that one would obtain from \eqref{FP-eq} a continuous of solutions parametrized by the initial conditions. In particular, denoting the initial conditions as\footnote{Obviously the choice of parametrization of the initial conditions is driven by the identification of the first two terms with the corresponding ones in the Einstein-Hilbert action, with $\Lt$ and $\tG$ the dimensionless cosmological and Newton's constants. The star denotes the fixed-point values of such parameters.}
\bea \label{a-g}
& \ft(0) = a_0 \equiv \f{\Lt^*}{8\pi \tG^*} \; , & \ft'(0) = a_1 \equiv -\f{1}{16\pi \tG^*} \; ,\;\;\;   \ft''(0) =a_2 \equiv \f{c^*}{(16\pi \tG^*)^2} \; ,
\eea
it would seem that the cosmological constant $\Lt$, the Newton's constant $\tG$ and the higher-derivative coupling $c^*$ are completely free at the fixed point.
However, an analysis similar to \cite{Morris:1994ki} should lead to the conclusion that only a finite subset of solutions satisfy basic regularity requirements, like not having singularities at finite $\Rt$.
Indeed we find that for generic initial conditions the solution develops a logarithmic singularity (see \eqref{mov-sing}) at a finite value of $\Rt$. We distinguish two type of singularities, movable singularities and fixed singularities. Singularities of the first kind appear because of the non-linear nature of the ODE, and as it is suggested by the name, they occur at a location that varies with the initial conditions. On the contrary, fixed singularities occur at those values of $\Rt$ where the equation is explicitly singular.
In the specific case, the latter correspond to the zeros of the denominator in \eqref{FP-eq}, which has three real zeros, at $\Rt=0$ and at $\Rt=\pm\sqrt{3 (9 + \sqrt{87})}\equiv \Rt_\pm$. The presence of such fixed singularities will play an important role in our analysis.

A brief explanation is due also for the fact that the equation \eqref{FP-eq} is third-order, given that in \eqref{FRGE} only the second functional derivative of $\G$ appears.
The reason is that, following a standard procedure \cite{Machado:2007ea,Codello:2008vh}, we adapt the cutoff $\cR_k$ to the Hessian of the action (see Sec.~\ref{Sec:cutoff}), meaning that the cutoff contains the second derivative $\ft_k''(\Rt)$. When the total derivative $d/dt$ acts on the cutoff in \eqref{FRGE}, we obtain a term $-2\Rt \ft_k'''(\Rt)$,  among others.
This explains at the same time why the equation is third-order, and why it has a singularity at $\Rt=0$: that is a point at which the coefficient of $\ft'''(\Rt)$ vanishes.
The appearance of such a fixed singularity is hence very generic, and it actually acts in favor of a qualitative scheme-independence: it is well known that a quantitative analysis of the FRGE depends on the cutoff choice, but one would expect that there would be even a qualitative difference between a third order (in the present scheme) and a second order (in an hypothetical scheme with no $\ft_k''(\Rt)$ dependence in the cuttoff) equation, i.e. a different number of initial conditions could lead to a continuos of fixed points or no fixed points in one or the other case. However, in presence of a singularity for the third-order equation the number of independent initial conditions at $\Rt=0$ is reduced to two by the requirement of regularity of the solution\footnote{A similar situation occurs in scalar field theories \cite{Comellas:1997tf} if as fundamental field one uses $y=\phi^2$ instead of $\phi$. In that case the regularity condition replaces the even potential condition $V'(\phi=0)=0$. We verified that the same could be done here, defining $U(x)=\ft(x^2)$, and the regularity condition being traded for the initial condition $U'(0)=0$. As nothing is gained in such scheme, while the gravitational interpretation gets a bit obscured, we stick in the following to the variable $\Rt$.}: upon substitution of a Taylor expansion $\ft(\Rt)=\sum_{n\geq 0} a_n \Rt^n$ we find that to lowest order in $\Rt$ the equation \eqref{FP-eq} imposes a condition,
\be \label{B0}
\begin{split}
B_- (a_0,a_1,a_2) \equiv\; & 24 a_0^2+17 a_1 a_0+216 a_2 a_0 \\
& +384 \pi ^2 \left(a_0+a_1\right) \left(2 a_0+3 a_1+18 a_2\right) a_0-27 a_1^2 -144 a_1 a_2 =0 \, ,
\end{split}
\ee
which for example can be easily solved for $a_2$.

The reason for the singularity at $\Rt_\pm$ is quite technical and will be explained in some detail in Sec.~\ref{Sec:Rt+}. Roughly speaking it has to do with the constant scalar mode.
We found its nature to be also quite generic, and we could not devise any reasonable scheme to make it disappear.
On the contrary, by a clever choice of cutoff scheme we were able to eliminate a number of other fixed singularities which appeared in previous versions of the equation \cite{Machado:2007ea,Codello:2008vh}.
Similarly to the singularity at the origin, we identify an analyticity condition at $\Rt_\pm$, equation \eqref{eqRt+}, relating the first three coefficients in a Taylor expansion of the solution at the singular point.

Unlike for finite $\Rt$, solutions are allowed to diverge at infinity. We find that a solution for $\Rt\to\infty$ must behave like $\ft(\Rt)\sim A \Rt^2$, with sub-leading corrections (detailed in Sec.\ref{Sec:infty}) dependending on the single free parameter $A$.
Such single parameter dependence provides an easier setting for numerical investigations (as compared to $\Rt=0$ where we have two free parameters), which can been performed by shooting backward from given initial conditions at large $\Rt$.

A true fixed point should correspond to a global solution, meaning a solution satisfying the analyticity conditions at the fixed singularities, presenting no other singularities at finite $\Rt$, and matching the large $\Rt$ expansion at infinity.
Due to the complexity of the equation, the quest for such a global solution turns out to be quite challenging, however with the help of preliminary results from numerical integrations we will show that these seemingly too many conditions can in principle be satisfied.

Postponing a comprehensive numerical investigation to future work, here we will exploit further the new insights that can be derived from the large $\Rt$ expansion.
In particular we find that treating such expansion as a standard truncation, we can identify an interesting structure for the fixed-point solutions (Fig.~\ref{fig:fixpmot}), and we can single out one point with surprising convergence properties. Unfortunately at such point the number of relevant directions seems to be increasing with the order of the truncation.

Finally, from the same large-$R$ asymptotic expansion, we conclude that if a global solution to the fixed-point equation exists, it must necessarily correspond to a fixed-point theory with effective action  $\G^*=A^* \int d^4 x \sqrt{g} R^2$, for some finite $A^*$.  
Such a result is also compatible with recent works \cite{Bonanno:2012jy,Hindmarsh:2012rc,Domazet:2012tw}, in which the identification $k^2\sim R$ is assumed for truncated versions of the average effective action.
Of course an $R^2$ action could also be expected on simple dimensional grounds, as it is the only $f(R)$ action possessing scale invariance, however it had never been verified before that this is indeed the case, as it is not obvious to see such scaling from truncated expansions, and it could in principle be not the case if the Ricci scalar had to acquire an anomalous dimension at the fixed point. Consistently with the LPA analogy we find here that the scaling is not anomalous.

The following sections will provide all the details for the interested reader.

\section{Hessian and gauge-fixing}
\label{Sec:Hessian}

In order to evaluate the r.h.s. of the FRGE, the first step is to compute the second variation of the ansatz at $h_{\m\n}=0$, i.e. the Hessian on the background.

Although functional variations of the action functional for  $f(R)$ theories have appeared before in the literature \cite{Machado:2007ea,Codello:2008vh}, we report them here again for completeness, and because of a slight difference in the choice of variables and notation.

We use the transverse-traceless decomposition of the metric fluctuations, given by
\be \label{TT-dec}
h_{\mu\nu} = h_{\mu\nu}^{T} + \na_\m \xi_\n + \na_\n \xi_\m + \na_\mu \na_\nu \s + \frac{1}{d} g_{\mu\nu} \hb\, ,
\ee
with the component fields satisfying
\be
g^{\mu \nu} \, h_{\mu\nu}^{T} = 0 \, , \quad \na^\mu h_{\mu\nu}^{T} = 0
\, , \quad \na^\mu \xi_\mu = 0 \, , \quad \hb = h +\D \s \, , \quad h=g_{\mu \nu} h^{\mu \nu} \, ,
\ee
and $\D=-\na^2$.

The Hessian for our ansatz comprises the following components:
\be
\Gb^{(2)}_{h^T_{\mu\nu}h^T_{\alpha\beta}}=
-\frac{Z_k}{2}\left[f_k'\left(\D+\frac{2}{d(d-1)}R\right)+\left(f_k-\f{2}{d}Rf_k'\right)\right]\delta^{\mu\nu,\alpha\beta}\ ,
\ee
\be
\Gb^{(2)}_{\xi_{\mu}\xi_{\nu}} =
\f{Z_k}{\a} \left(\D-\frac{R}{d}\right)\left[\left(\D-\frac{R}{d}\right)+ \a\left( \frac{2R}{d}f_k'-f_k\right)\right]
g^{\mu\nu} \, ,
\ee
\be
\Gb^{(2)}_{\hb\hb}
=  Z_k\frac{d-2}{2d^2}\left[ \f{2(d-1)^2}{d-2} f_k''\left(\D-\frac{R}{d-1}\right)^2+(d-1) f_k'\left(\D-\frac{R}{d-1}\right)
-\left(Rf_k'-\f{d}{2}f_k\right) \right]  \, ,
\ee
\be
\Gb^{(2)}_{\hb\sigma}
= Z_k\frac{d-2}{2d^2}\left(Rf_k'-\f{d}{2}f_k\right)  \D \, ,
\ee
\be
\Gb^{(2)}_{\sigma\sigma}
=  
Z_k\frac{d-2}{2d^2 \a} \left[\f{2(d-1)^2}{d-2}\left(\D-\frac{R}{d-1}\right)
+\a\frac{d}{d-2}\left(Rf_k'-\f{d}{2}f_k\right)  
\right]\left(\D-\frac{R}{d-1}\right) \D \ .
\ee
We have already included in the Hessian the terms coming from the gauge-fixing action
\be
\G_{\rm gf}[h, g] = \f{Z_k}{2\a} \int d^d x\sqrt{g} F_\m F^\m \, ,
\ee
with
\be
F_\m \equiv \na^\n h_{\m\n} - \f{1}{d} \na_\m h  = -\Big(\D -\f{R}{d}\Big) \xi_\m -\na_\m \Big(\f{d-1}{d}\D-\f{R}{d}\Big)\s \, .
\ee

In the gauge $\a\to 0$, that we will use in the following, neither $\Gamma^{(2)}_{\hb\sigma}$ nor the other terms proportional to the equations of motion in $\Gamma^{(2)}_{\s\s}$ and $\Gamma^{(2)}_{\xi\xi}$ contribute to the traces.

\subsection{Ghosts and auxiliary fields}
\label{Sec:ghosts}

For the ghost action we follow \cite{Benedetti:2011ct} and write
\be \label{ghosts2}
\begin{split}
\G_{\rm gh} = \f{Z_k}{\a} \int d^d x \sqrt{g} \Big\{ & \Cb^{T\m} \Big( \D -\f{R}{d}\Big)^2 C_\m^T +4\Big( \f{d-1}{d}\Big)^2 \cb \Big( \D -\f{R}{d-1}\Big)^2 \D c    \\
   & +  B^{T\, \m} \Big( \D -\f{R}{d}\Big)^2 B_\m^T +4 \Big( \f{d-1}{d}\Big)^2 b \Big( \D -\f{R}{d-1}\Big)^2 \D b \Big\}\, ,
\end{split}
\ee
where the $C_\m^T$ and $c$ are complex Grassmann fields, while $B_\m^T$ and $b$ are real fields, and the index $T$ denotes transverse vectors.
As explained in \cite{Benedetti:2011ct}, while being formally equivalent to the standard implementation, this version of the ghost sector has the merit of realizing
an exact cancellation (generally on shell, but also off shell in the $\a= 0$ gauge) between ghosts and pure-gauge degrees of freedom, thus ensuring gauge-independence of the on-shell effective action. We have explicitly verified that all the qualitative features emerging in the following analysis are found also with a standard version of the ghost action.

The final ingredient of our truncation is the action for the auxiliary fields, introduced to take into account the Jacobian arising in the TT decomposition \eqref{TT-dec}.
The Jacobian for the gravitational sector leads to the auxiliary action
\be \label{aux-gr}
\begin{split}
S_{\rm aux-gr} =  \int d^d x \sqrt{g} \Big\{ & 2 \bar{\chi}^{T\, \m} \Big( \D -\f{R}{d}\Big) \chi_\m^T + \Big( \f{d-1}{d}\Big) \bar{\chi} \Big( \D -\f{R}{d-1}\Big) \D \chi    \\
   & + 2 \z^{T\m} \Big( \D -\f{R}{d}\Big) \z_\m^T +\Big( \f{d-1}{d}\Big) \z \Big( \D -\f{R}{d-1}\Big) \D \z  \Big\}\, ,
\end{split}
\ee
where the $\chi_\m^T$ and $\chi$ are complex Grassmann fields, while $\z_\m^T$ and $\z$ are real fields.
The Jacobian for the transverse decomposition of the ghost action is given by
\be
S_{\rm aux-gh} = \int d^d x \sqrt{g} \, \phi \D \phi \, ,
\ee
with $\phi$ a real scalar field.

\section{Cutoff scheme}
\label{Sec:cutoff}

We will use here a variation of the scheme introduced in  \cite{Benedetti:2011ct}, which we will call ``on-shell'' Type II cutoff (in the spirit of the nomenclature of \cite{Codello:2008vh}).

Denoting with $r_k(z)$ some fixed cutoff profile function, we recall that the scheme introduced in \cite{Benedetti:2011ct} amounts to choosing the cutoff $\cR_k$ in such a way to implement in the on-shell part of Hessian the rule
\be \label{cutrule}
\D\to P_k\Big( \f{\D}{k^2} \Big) \equiv \D+k^2 r_k\Big( \f{\D}{k^2} \Big) \, .
\ee
In the course of the present work we realized that certain unphysical singularities appearing in the functional traces (see \cite{Benedetti:2011ct}, but also  \cite{Machado:2007ea} and \cite{Codello:2008vh}) actually appear from this unfortunate choice. Consider for example the typical operators appearing in the scalar part, $\D_0 = \D - \f{R}{d-1}$.
When using \eqref{cutrule} in combination with the optimized cutoff \eqref{optcutoff}, we obtain for the modes below $k^2$,
\be
\D_0 \to k^2 -  \f{R}{d-1} \, ,
\ee
which is of course zero, and hence not invertible, at $\Rt = d-1$.
In order to avoid such singularities, we will adopt here the following set of replacement rules:
\bea \label{newrule0}
\D_0 \equiv \D - \f{R}{d-1} &\to &   P_k^{(0)} \Big( \f{\D_0}{k^2} \Big) \equiv \D_0 + k^2 r_k\Big( \f{\D_0}{k^2} \Big) \, , \\
\label{newrule1}
\D_1 \equiv \D - \f{R}{d} &\to &   P_k^{(1)} \Big( \f{\D_1}{k^2} \Big) \equiv \D_1 + k^2 r_k\Big( \f{\D_1}{k^2} \Big) \, , \\
\label{newrule2}
\D_2 \equiv \D + \f{2R}{d(d-1)} &\to &   P_k^{(2)} \Big( \f{\D_2}{k^2} \Big) \equiv \D_2 + k^2 r_k\Big( \f{\D_2}{k^2} \Big) \, ,
\eea
and the profile function $r_k(z)$ will be chosen later.

Note that as we will work in the gauge $\a=0$ the difference between on-shell and off-shell scheme is not essential, and it is basically only used as a motivation for including in the $\G^{(2)}_{h^T_{\m\n}h^T_{\a\b}}$ and $\G^{(2)}_{\hb\hb}$ operators certain potential terms and not others.\footnote{We recall that the crucial aspect of the ``on-shell type'' cutoff introduced in   \cite{Benedetti:2011ct} was to avoid introducing a gauge-breaking cutoff term, that is a cutoff term for the gauge-variant field components $\xi_\m$ and $\s$ with no ghost counterpart. In the gauge $\a=0$ the fields  $\xi_\m$ and $\s$ only survive in the gauge-fixing part of the Hessian, and the effect of the regulator on such terms is correctly taken into account by \eqref{ghosts2} in combination with (\ref{newrule0}-\ref{newrule1}).}

In practice, we have the following cutoff functions
\be
\cR_k^{h^T_{\m\n}h^T_{\a\b}} = -\frac{Z_k}{2} f_k'\, k^2 r_k\left(\f{\D_2}{k^2}\right)\delta^{\mu\nu,\alpha\beta} \, ,
\ee
\be
\cR_k^{\xi_{\mu}\xi_{\nu}} =
\f{Z_k}{\a} \left[ \left(P_k^{(1)} \Big( \f{\D_1}{k^2} \Big)\right)^2- \D_1^2    \right]
g^{\mu\nu} \, ,
\ee
\be \label{cutoff-hh}
\cR_k^{\hb\hb}
=   Z_k\frac{d-2}{2d^2}\left[ \f{2(d-1)^2}{d-2} f_k''\left(\left(P_k^{(0)} \Big( \f{\D_0}{k^2} \Big)\right)^2-\D_0^2\right)  
 +(d-1) f_k'  k^2 r_k\Big( \f{\D_0}{k^2} \Big) \right] \, ,
\ee
\be
\cR_k^{\s\s}
=   Z_k\frac{(d-1)^2}{d^2 \a}  \left(\left(P_k^{(0)} \Big( \f{\D_0}{k^2} \Big)\right)^3 + \f{R}{d-1} \left(P_k^{(0)} \Big( \f{\D_0}{k^2} \Big)\right)^2 -\D_0^2 (\D_0+\f{R}{d-1}) \right) \, ,
\ee
and so on for ghosts and auxiliary fields.

\section{Spectral sums}
\label{Sec:sums}

We are going to evaluate the traces by a direct spectral sum rather than in a heat kernel expansion.
By spectral sum we mean that a generic trace will be evaluated as
\be \label{eigsum}
\Tr_s W(\D) = \sum_n D_{n,s} W(\l_{n,s}) \, ,
\ee
where $\{\l_{n,s}\}$ is the spectrum of eigenvalues of the Laplace-type operators (\ref{newrule0}-\ref{newrule2}) on spin-$s$ fields, with the relative multiplicities $\{D_{n,s}\}$.
The spectra and multiplicities are reported in Table~\ref{table}, and are obtained by appropriately shifting the standard one (for example see  \cite{Rubin:1983be}).
\begin{table}
\begin{center}
\begin{tabular}{|c|c|c|}\hline
Spin s  & Eigenvalue $\l_{n,s}$ & Multiplicity $D_{n,s}$\\\hline
0 & $\frac{n(n+d-1)-d}{d(d-1)}R$; $n=0,1\ldots$& $\frac{(n+d-2)!\, (2n+d-1)}{n!(d-1)!}$\\\hline
1 & $\frac{n(n+d-1)-d}{d(d-1)}R$; $n=1,2\ldots$& $\frac{(n+d-3)!\, n(n+d-1)(2n+d-1)}{(d-2)!(n+1)!}$\\\hline
2 & $\frac{n(n+d-1)}{d(d-1)}R$; $n=2,3\ldots$& $\frac{(n+d-3)!\, (d+1)(d-2)(n+d)(n-1)(2n+d-1)}{2(d-1)!(n+1)!}$\\\hline
\end{tabular}\end{center}
\caption{Eigenvalues of the Laplace-type operators (\ref{newrule0}-\ref{newrule2}) on the $d$-sphere and their multiplicities}
\label{table}
\end{table}

We have to be careful not to include fictitious modes in the sum. 
Remembering our decomposition for the metric fluctuations \eqref{TT-dec},
we see that we should exclude two sets of modes that give no contribution to $h_{\m\n}$. First, we should exclude the Killing vectors, satisfying $\na_\m \xi_\n + \na_\n \xi_\m=0$. Second, we should leave out also the constant scalar modes $\s =$ constant.
A similar set of modes should be excluded also from the ghosts and auxiliary fields, as these are all fields introduced hand-in-hand with $\xi$ and $\s$. The only fields for which we retain all the modes are $h^T_{\m\n}$ and $\hb$.
Note that, differently from \cite{Codello:2008vh,Machado:2007ea}, we do not exclude the scalar modes corresponding to conformal Killing vectors $\cC_\m= \na_\m\s$, $i.e.$ those scalar modes satisfying $ \na_\mu \na_\nu \s =  \frac{1}{d} g_{\mu\nu} \na^2 \s$. It is indeed clear that in our decomposition \eqref{TT-dec} such modes do contribute to $h_{\m\n}$. 
This can be seen also from the point of view of the ghosts: the ghost modes should be in one-to-one correspondence with the modes of the gauge parameter $(\eps_\m^T,\eps)$, and from $\cL_\eps g_{\m\n} = \na_\m \eps^T_\n + \na_\n \eps^T_\m + 2 \na_\mu \na_\nu \eps$ it is obvious that there is no reason to exclude the scalar modes $\eps$ corresponding to conformal Killing vectors.
As a consequence the tensor and vector sums will start at $n=2$, while all the scalars sums will begin at $n=1$, except for the $\hb$ mode starting at $n=0$.

We choose to work with Litim's optimized cutoff \cite{Litim:2001up}
\be \label{optcutoff}
r_k(z) = (1-z) \th(1-z) \, ,
\ee
for which
\be
\p_t (k^2 r_k(\D/k^2) ) = 2 k^2 \th (k^2-\D) \, .
\ee
Its great technical advantage is that with it all the functions appearing in the FRGE have a numerator proportional to the step function, and hence the spectral sums are cut off at 
$N_s=\text{max} \{n\in\mathbb{N} : \l_{n,s}\leq k^2\}$. At the same time, for all $\l_{n,s}\leq k^2$, we have $P_k(\l_{n,s}/k^2) = k^2$.

We write the FRGE as
\be \label{sumFRGE}
\begin{split}
\p_t \Gb_k &=  \sum_{n=2}^{N_2(\Rt)} W_2(\l_{n,2}/k^2, \Rt)+\sum_{n=2}^{N_1(\Rt)} W_1(\l_{n,1}/k^2, \Rt)\\
     & \qquad +\sum_{n=1}^{N_0(\Rt)} W_0^{\text{np}}(\l_{n,0}/k^2, \Rt)+\sum_{n=0}^{N_0(\Rt)} W_0^{\hb}(\l_{n,2}/k^2, \Rt) \\
   &\equiv  \cT_2 + \cT_1 +\cT^{\text{np}}_0+\cT^{\hb}_0 \, .
\end{split}
\ee
where the functions $W_s(\D/k^2, \Rt)$ are obtained by collecting the contributions to \eqref{FRGE} coming from all the fields of spin $s$, and we have separated the $\hb$ contribution from that of the other scalars (dubbed ``np'', for non-physical).
Note that $N_s$ is a function of $\Rt$ as well as of the spin $s$.

Explicitly, the functions $W_s(z,\Rt)$ in $d=4$ are given by
\be \label{W2}
W_2(z,\Rt) = \frac{(z-1) \left(\p_t\ft'(\Rt)-2 \Rt \ft''(\Rt)+2
   \ft'(\Rt)\right)-2\ft'(\Rt)}{(\Rt-2) \ft'(\Rt)-2
   \ft(\Rt)} \, ,
\ee
\be \label{W1}
W_1(z,\Rt) = W_0^{\text{np}}(z,\Rt) = -1 \, ,
\ee
\be \label{W0}
\begin{split}
W_0^{\hb}(z,\Rt) = & \frac{3 (1-z^2) \left(3 \p_t\ft''(\Rt)-6 \Rt   \ft^{(3)}(\Rt)\right)}{18 \ft''(\Rt)-2 (\Rt-3)   \ft'(\Rt)+4 \ft(\Rt)}
 \\   &   +\frac{3 (1- z )  \left(\p_t\ft'(\Rt)-2 \Rt \ft''(\Rt)+2   \ft'(\Rt)\right)}{18 \ft''(\Rt)-2 (\Rt-3)   \ft'(\Rt)+4 \ft(\Rt)} \, .
\end{split}
\ee
Note the simple form of $W_1(z,\Rt)$ and $W_0^{\text{np}}(z,\Rt)$, which is due to the ghost choice \eqref{ghosts2}. With a standard ghost action we find slightly more complicated expressions, in particular with a dependence on $\eta=-\p_t\log Z$, and for the fixed-point equation we have to require also 
$\eta=-2$, beside $\p_t\ft(\Rt)=0$. Once that is done, all the qualitative features of our analysis are left invariant.

\subsection{Approximating the sums}
\label{Sec:SumApprox}

Thanks to the cutoff choice \eqref{optcutoff}, all the sums appearing in the FRGE can be performed analyti-cally. Indeed they all involve simple sums of the following type
\be
\cS_{m,s}(\Rt) = \sum_{n=n_{s}}^{N_s(\Rt)} D_{n,s}\, \left(\f{\l_{n,s}}{k^2}\right)^m \, ,
\ee
for $m\in\{0,1,2\}$.

Unfortunately, the use of \eqref{optcutoff} is not safe from pitfalls. The main drawback is that since the upper bound on the summation, $N_s(\Rt)$, is a staircase-function of 
the curvature $\Rt$, also the resulting sum $\cS_{m,s}(\Rt)$ is a staircase-function.
Dealing with differential equations containing staircase-functions is quite unpleasant, and for this reason we will adopt some smoothing strategy. Of course such smoothing constitutes an approximation and it introduces an additional scheme dependence. We have explicitly verified that all the qualitative conclusions of our work are left unaltered by use of different smoothing choices.

We now illustrate some possible smoothing choices, by means of the simplest example, that is, the function $\cS_{0,0}(\Rt)$ with $n_0=0$, also known as the spectral counting function.
Specializing to $d=4$ dimensions, we easily find
\be
\cS_{0,0}(\Rt)=\cP(\lfloor N_0(\Rt)\rfloor )
\ee
where
\be
\cP(N)=\sum_{n=0}^{N} D_{n,0} = \f{1}{12} (1 + N) (2 + N)^2 (3 + N) \, ,
\ee
and
\be
N_0(\Rt) =  \f{-3 \Rt + \sqrt{\Rt (48 + 25 \Rt)}}{2 \Rt} \, ,
\ee
and where $\lfloor . \rfloor$ denotes the floor function, which is what gives rise to the staircase nature of the function.

In \cite{Benedetti:2011ct} the simplest approximation was made, replacing $\lfloor x \rfloor\to x$,
that is, defining
\be
\cS^{(+)}_{0,0}(\Rt)=\cP(N_0(\Rt) ) \, .
\ee
\begin{figure}[t]
\centering \includegraphics[width=8cm]{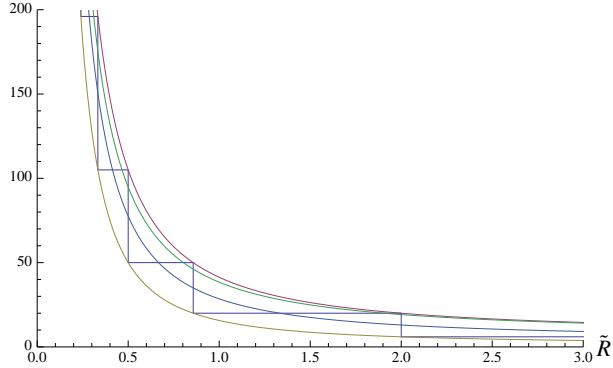}
  \caption{A plot of the staircase function $\cS_{0,0}(\Rt)$, together with four possible smoothing-curves, from top to bottom: the upper-edge, the asymptotic, the mean and the lower-edge interpolations.}
\label{fig:stairs}
\end{figure}
This is the smoothing which we will call ``upper-edge'', as it touches the original staircase function on all the upper edges of its steps.
Analogously, we can define
\be
\cS^{(-)}_{0,0}(\Rt)=\cP(N_0(\Rt)-1 ) \, ,
\ee
which we will call ``lower-edge'', as it touches the lower edges of the steps (see Fig.~\ref{fig:stairs}).

Such smoothing interpolations have the obvious disadvantage of containing square roots, which also are not particularly welcome in the differential equation.
Rational interpolating functions can be obtained in several ways.
One, is to consider the average of the upper- and lower-edge interpolations, which we will call ``mean'' interpolation:
\be
\cS^{(\text{mean})}_{0,0}(\Rt) = \f12 \left( \cS^{(+)}_{0,0}(\Rt)+\cS^{(-)}_{0,0}(\Rt)\right) \, .
\ee

We can also introduce an interpolation just by requiring that it matches the asymptotic leading behavior of the original function at $\Rt\to 0$ and at $\Rt\to+\infty$
(a similar choice was used in \cite{Reuter:2008qx}).
We call such function the ``asymptotic'' interpolation, and in the specific example it reads
\be
\cS^{(\text{asymp})}_{0,0}(\Rt) = \f{12}{\Rt^2} + 1 \, .
\ee

Finally, one could use the heat kernel interpolation, as tacitly done in \cite{Machado:2007ea,Codello:2008vh}.
On the relation between counting functions and heat kernel there is abundant literature (see for example \cite{Dai:2009zza} and references therein),
and we will not indulge on that here.

We found no evident qualitative difference between the mean, the asymptotic and the heat kernel interpolations, and we will report in the following only the results obtained within the asymptotic interpolation.

\section{The fixed-point differential equation}
\label{Sec:FPeq}

The full FRGE in $d=4$ dimensions takes the form
\be \label{evalFRGE}
\f{384\pi^2}{\Rt^2} \left( \p_t \ft_k(\Rt) + 4 \ft_k(\Rt) - 2 \Rt \ft_k'(\Rt)    \right) = \cT_2 + \cT_1 +\cT^{\text{np}}_0+\cT^{\hb}_0 \, ,
\ee
where we used \eqref{lhs} and the explicit formula for the volume of $S^4$, and where
we have subdivided the r.h.s. into the contributions of the TT-tensor modes
\be
\cT_2 = -\frac{20 \left(\p_t\ft'(\Rt)-2 \Rt \ft''(\Rt)+8
   \ft'(\Rt)\right)}{\Rt^2 \left((\Rt-2) \ft'(\Rt)-2
   \ft(\Rt)\right)} \, ,
\ee
the vector modes
\be
\cT_1 = -\f{36}{\Rt^2} \, ,
\ee
the non-physical scalar modes (by which we mean all the scalars but the trace $\hb$)
\be
\cT^{\text{np}}_0 = - \f{12+5 \Rt^2}{\Rt^2} \, ,
\ee
and finally the contribution of the trace mode $\hb$
\be \label{Th}
\begin{split}
\cT^{\hb}_0 = & \frac{ 1}{
   2 \Rt^2 \left(-9 \ft''(\Rt)+(\Rt-3) \ft'(\Rt)-2
   \ft(\Rt)\right)
   } \times \\
&\quad \Big\{ \left(\Rt^4-54 \Rt^2-54\right) \left( \p_t\ft''(\Rt) -2\Rt  \ft^{(3)}(\Rt) \right) \\
&\qquad  -\left(\Rt^3+18 \Rt^2+12\right) \left(  \p_t\ft'(\Rt)     -2\Rt \ft''(\Rt) + 2    \ft'(\Rt) \right) \Big\} \, .
\end{split}
\ee
An analysis of such a partial differential equation seems a formidable task, and we will limit ourselves just to the associated fixed-point ordinary differential equation.

At the fixed point $\p_t\ft(\Rt)=0$, and we can express the ODE in canonical form, by solving for $\ft^{(3)}(\Rt)$, resulting in
\be \label{FP-eq2}
\ft^{(3)}(\Rt) = \frac{\cN( \ft,\ft',\ft''; \Rt)}{ \Rt (\Rt^4-54 \Rt^2-54) \left(( \Rt-2 ) \ft'(\Rt) -2 \ft(\Rt)\right)} \, ,
\ee
where
\be \label{eq-numerator}
\begin{split}
\cN( \ft,\ft',\ft''; \Rt) = & -360 \Rt \ft''(\Rt)^2+768 \pi ^2 \Rt \left(\Rt^2-5
   \Rt+6\right) \ft'(\Rt)^3 \\
& +\ft(\Rt)^2 \left(-27648 \pi ^2
   \ft''(\Rt)+3072 \pi ^2 (3 \Rt-5) \ft'(\Rt)-20
   \Rt^2-192\right) \\
& +\ft(\Rt) \left(-2 \left(\Rt^4+18 \Rt^3+45
   \Rt^2+52 \Rt+432\right) \ft''(\Rt) \right.\\
& \qquad  -1536 \pi ^2 \left(3
   \Rt^2-10 \Rt+6\right) \ft'(\Rt)^2 \\
& \qquad \left. +\ft'(\Rt) \left(27648
   \pi ^2 (\Rt-1) \ft''(\Rt)+22 \Rt^3-14 \Rt^2+192
   \Rt-136\right)\right) \\
& +\ft'(\Rt)^2 \left(-6912 \pi ^2 (\Rt-2)
   \Rt \ft''(\Rt)-6 \Rt^4+9 \Rt^3-42 \Rt^2+68
   \Rt+216\right) \\
& +\left(\Rt^5+16 \Rt^4+9 \Rt^3-38 \Rt^2+288
   \Rt+576\right) \ft'(\Rt) \ft''(\Rt)-6144 \pi ^2
   \ft(\Rt)^3 \, .
\end{split}
\ee
%

\subsection{Gaussian fixed point}
\label{Sec:gfp}

The Gaussian fixed-point solution is easily recovered as in \cite{Machado:2007ea}. We rescale $\ft(\Rt)\to \f{1}{\b} \ft(\Rt)$ and look for a solution in the limit $\b\to 0$.
The r.h.s. of \eqref{evalFRGE} is homogeneous of degree zero in $\ft$ and its derivatives, while the l.h.s. is of degree one, hence in the limit of vanishing coupling we get
the fixed-point equation
\be \label{gfpFRGE}
 2 \ft(\Rt) - \Rt \ft'(\Rt)     = 0 \, ,
\ee
whose unique solution is $\ft(\Rt) = a \Rt^2$, for some constant of integration $a$. Of course the ``Gaussian'' interpretation is as usual: expanding the metric as $g_{\m\n}=\gb_{\m\n}+\sqrt{\b} H_{\m\n}$, where the background $\gb_{\m\n}$ is a solution of the equations of motion, the only part surviving in the $\b\to 0$ limit is the one quadratic in $H_{\m\n}$.

In the following section we will investigate the possible existence of a non-Gaussian fixed-point solution of \eqref{FP-eq2}.

\section{Analysis of the fixed-point equation}
\label{Sec:analysis}

We are now going to discuss the following properties of the equation \eqref{FP-eq2}: fixed singularities at $\Rt=0$ and $\Rt=\Rt_+$; movable singularities; asymptotic behavior at infinity. Performing numerical integrations and putting things together, we will draw some conclusions about the existence of non-trivial global solutions.

\subsection{Singularity at $\Rt=0$, and polynomial truncations}
\label{Sec:poly}

As anticipated in Sec.~\ref{Sec:general}, the fixed singularity at the origin is there for the same reason why the equation is of third order, i.e. because because $f''(R)$ appears in the cutoff function \eqref{cutoff-hh} and because of the dimensional nature of $R$.

The presence of such a singularity acts in a way to reduce the number of independent initial conditions.
This can be easily understood: a regular solution of an equation of the canonical type $\ft'''(\Rt)=F(\ft,\ft',\ft'';\Rt)$ can be constructed in the neighborhood of any regular point $\Rt_0$, substituting a series expansion $\ft(\Rt)=\sum_{n\geq 0}  a_n (\Rt-\Rt_0)^n$, expanding in series the resulting $F$, and then imposing the equation order by order. 
The first of such equations will fix $a_3$ as a function of $a_0$, $a_1$ and $a_2$, the next will fix $a_4$, and so on, leaving the first three series coefficients free.
However if $F$ has a pole of order $n$ at $\Rt_0$, the r.h.s. will produce a Laurent series with $n$ singular terms not matched on the l.h.s. (analytic by construction), and which have to be equated to zero; the first of such equations will only contain the lowest order expansions for $\ft$, $\ft'$ and $\ft''$, hence it will only depend on $a_0$, $a_1$ and $a_2$, thus providing a constraint on the initial conditions. This is precisely what happens at $\Rt_0=0$, where we have a simple pole in \eqref{FP-eq2}, 
\be \label{serie0}
0 = \f{2 B_-(a_0,a_1,a_2)}{27 (a_0+a_1) \Rt} + \sum_{n\geq 0} B_n(a_0,...,a_{n+3}) \Rt^n \, ,
\ee
leading to the constraint \eqref{B0}.
At higher orders, $a_{n+3}$ enters linearly in $B_n$, hence the tower of equations obtained from \eqref{serie0} can be solved iteratively, providing a series expansion of the solution, as function of the two initial data $a_0$ and $a_1$.

The usefulness of the series solutions for the search of a global solution ends here, as in general it will only be valid at small $\Rt$. In particular the convergence radius of the solution is generically limited by the presence of singularities in the complex plane, rendering it difficult to search for initial conditions leading to a regular solution on the whole real axis. 
We do not report here results obtained using either ratio or root test, as they performed very poorly.
A more sophisticated analysis using Pad\'e approximants was also attempted, but the presence of two initial conditions variables rather than one makes a systematic study a bit clumsy, and we gained from it no useful insights to report here.

The polynomial truncations studied so far in the literature can be understood as an approximate way to reduce the $\infty^2$ set of solutions to a finite subset by discarding the contribution of all the terms in the series expansion beyond a certain order $N$. In other words, one makes the ansatz
\be \label{truncation}
\ft(\Rt)=\sum_{n=0}^N  a_n \Rt^n \, ,
\ee
and then imposes the first $N+1$ equations coming from \eqref{serie0}, that is, two more than we would impose to solve at order $N$ the Cauchy problem for \eqref{FP-eq2} with initial conditions $\ft(0)=a_0$ and $\ft'(0)=a_1$. 
Yet another way to say it is that we solve \eqref{serie0} up to $a_{N+2}$ (that is, up to $B_{N-1}=0$), and then impose $a_{N+1}=a_{N+2}=0$.
From the point of view of the differential equation, the imposition of such condition can be understood as an heuristic way to push farther away singularities \cite{Morris:1994ki}.

Proceeding on such a route we find similar results to previous analyses. For example, truncating at $N=1$ we recover the Einstein-Hilbert truncation, and we find
a non-Gaussian fixed point at $a_0=0.005984$, $a_1= -0.01497$, corresponding to $\Lt^*=0.1998$ and $\tG^*=1.329$, with critical exponents
$\th_\pm=1.62425 \pm i 3.42642$.
Actually one finds a number of fixed-point solutions which increase with $N$, but most of them are there for one truncation and not for another, or they are at unacceptable values (complex couplings, or negative Newton's constant). These are generally considered spurious fixed points and hence discarded, however a precise and reliable method to select which fixed points to keep and which to discard in this scheme is missing.
In particular a systematic expansion and analysis of the convergence with $N$ is very demanding from the computational point of view, essentially due the fact that the series solution of \eqref{serie0} leads to high-order polynomials of two variables for the coefficients $a_{N+1}(a_0,a_1)$ and $a_{N+2}(a_0,a_1)$.

\subsection{Singularity at $\Rt=\Rt_+$}
\label{Sec:Rt+}

At $\Rt=\Rt_+\equiv \sqrt{3 (9 + \sqrt{87})} \simeq 7.415$ the equation has another fixed singularity in the form of a simple pole.
The origin of such a singularity has to be looked for in the $\hb\hb$ sector of the theory. With the operator choice in \eqref{newrule0} we have eliminated any singularity in the equation \eqref{evalFRGE}, 
but we have shifted the eigenvalue of the constant $\hb$ mode to a negative value: from Table~\ref{table} we have $\l_{0,0}=-R/3$.
Being negative, such mode is never excluded from the sum (the step function in the optimized cutoff only kills modes larger than $k^2$, which of course is positive). 
The $\ft'''(\Rt)$ term in \eqref{W0} is multiplied by a factor proportional to $\sum_{n\geq 0} (1-\l_{n,0}^2/k^4)$, and due to the lowest mode this can become zero at some large value of $R$, where thus the equation develops a singularity. This is precisely what happens at $\Rt_+$.

As a result of the pole, if we Taylor expand the solution around $\Rt_+$ we find again a condition that reduces the number of independent initial conditions to two.
We write the expansion as
\be \label{serieR+}
\ft(\Rt)=b_0 +b_1 (\Rt-\Rt_+) +\f{b_2}{2} (\Rt-\Rt_+)^2  + \sum_{n\geq 3} b_n (\Rt-\Rt_+)^n \, ,
\ee
To leading order the equation \eqref{FP-eq2} reduces to
\be \label{eqRt+}
0 =  \f{  \tilde{B}_+(b_0,b_1,b_2) }{ \Rt-\Rt_+ } +O\Big( (\Rt-\Rt_+)^0 \Big) \, , 
\ee
where
\be \label{Bt+}
\tilde{B}_+(b_0,b_1,b_2)= \f{  B_+(b_0,b_1,b_2) }{ 12 \sqrt{87} \Rt_+^2  ( b_1( \Rt_+ -2  ) -2 b_0  ) } \, ,
\ee
and
\be \label{B+}
\begin{split}
B_+(b_0,b_1,b_2)= &   -768 \pi ^2 b_1^3 \Rt_+ \left(\Rt_+^2-5 \Rt_++6\right)+b_1^2\left( -9 \Rt_+^3+366 \Rt_+^2-68 \Rt_++108\right)\\
&   + 768 \pi ^2 b_1^2 \left((9 b_2+6 b_0 )  \Rt_+^2-2 (9 b_2+10 b_0 ) \Rt_+ +12 b_0 \right)  \\
&   -b_1 b_2 \left(27648 \pi ^2 b_0   (\Rt_+-1)+63 \Rt_+^3+826 \Rt_+^2+342 \Rt_++1440\right)\\
& - 2 b_0 b_1  \left(1536 \pi^2 b_0 (3 \Rt_+-5)+11 \Rt_+^3-7 \Rt_+^2+96 \Rt_+-68\right)\\
 & +2 b_0 b_2 \left(13824 \pi ^2 b_0+18 \Rt_+^3+99 \Rt_+^2+52   \Rt_++486\right) \\
 & + b_0^2 \left(6144 \pi ^2 b_0 +20 \Rt_+^2+192\right) +360 b_2^2 \Rt_+ \, .
\end{split}
\ee
In this case we find that the regularity condition is $\tilde{B}_+(b_0,b_1,b_2)=0$.

\subsection{Movable singularities}
\label{Sec:movsing}

It is well known that nonlinear ODEs can have also movable singularities beside the fixed ones.

For our equation, we found that for generic initial conditions the solution ends at some $\Rt=\Rt_c$ where the solution becomes singular. 
More precisely singular solutions can be constructed at any point $\Rt=\Rt_c\neq 2$, of the form
\be \label{mov-sing}
\ft(\Rt) \sim \log |\Rt-\Rt_c | \left(m_0 +m_1 (\Rt-\Rt_c) + O\left( (\Rt-\Rt_c)^2\right) \right) +c_0 +c_1 (\Rt-\Rt_c) + O\left( (\Rt-\Rt_c)^2\right) \, ,
\ee
where the only free parameters are $c_0$ and $\Rt_c$, all the others being determined as function of those. For example,
\be
m_0 = \frac{\Rt_c^5-2 \Rt_c^4-54 \Rt_c^3+108 \Rt_c^2-54 \Rt_c+288}{3456\, \pi^2 (\Rt_c-2)} \, .
\ee
%

\subsection{Expansion at $\Rt\to\infty$}
\label{Sec:infty}

The solution can also be worked out in the limit of $\Rt\to\infty$. This actually turns out to be the limit in which the series expansion is most under control.

\begin{figure}[ht]
\centering \includegraphics[width=9cm]{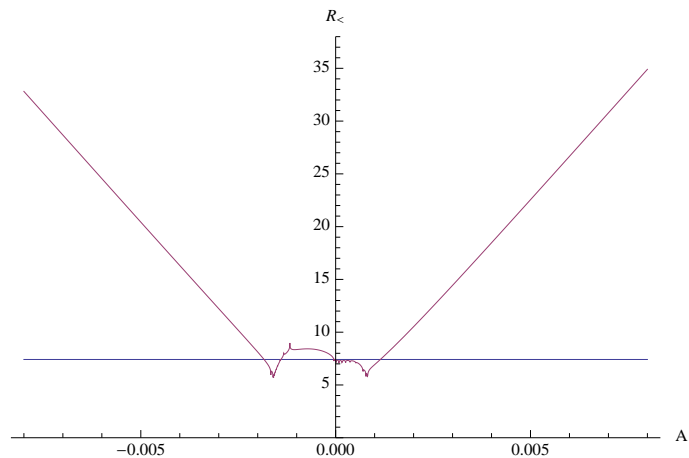}
\caption{\label{fig:seriesingA}
\small{A plot of the (inverse) radius of convergence $\Rt_<$ for the series in \eqref{asymp-sol} as a function of $A$ (the series converges for $\Rt>\Rt_<$).
The horizontal line is $\Rt=\Rt_+$.}
}
\end{figure}

We found that the asymptotic solution has an expansion of the following form:
\be \label{asymp-sol}
\ft(\Rt) \sim A \Rt^2 \left( 1 +\sum_{n\geq 1} d_n \Rt^{-n} \right) \, .
\ee
All the coefficients $d_n$ can be fully determined iteratively as function of only $A$, which is another crucial point for our results. 
The presence of only one free parameter in the asymptotic expansion is a very general property of the equation, deriving from the higher order nature of the pole at infinity (as perhaps more easily seen after mapping the point at infinity to a finite value, and studying the balance of terms in \eqref{FP-eq2}).
We report here the expression for the first three terms of the asymptotic solution:
\be
d_1=-\f{15}{2}\, , \;\;\; d_2= -\f94 (5 + 384 A \pi^2)\, , \;\;\; d_3= -\f37 (-17 + 17640 A \pi^2 + 1161216 A^2 \pi^4) \, .
\ee
We have computed coefficients up to $d_{30}$, and we found that in general the coefficient $d_n$ is a polynomial of order $n-1$ in $A$, with the coefficient of the highest power a number of order $10^{3n-3}$,
suggesting that, at least for large enough $A$, the series should be convergent for $\Rt\geq \Rt_< \sim 10^3 A$.
We have verified this conclusion by the root test (or Cauchy--Hadamard theorem), i.e. by a linear fit\footnote{We have checked that in the explored range of values for $A$ the dependence of $r_n$ on $1/n$ is quite close to linear, only with a slight convexity at certain small values of $A$. For comparison a ratio method was attempted too, i.e. taking $r_n=|d_{n+1}/d_n|$, but the resulting behavior seemed far from linear, and for this reason we stick here to the root test results.} 
of $r_n=\sqrt[n]{d_n}$ as function of $1/n$ to estimate $\Rt_<=\lim_{n\to\infty} r_n$.
A plot of $\Rt_<$ as function of $A$ is reported in Fig.~\ref{fig:seriesingA}.
As expected, for sufficiently large values of $A$, $R_<$ grows linearly with $A$. At small $A$ some spikes are observed in correspondence of the zeros of the coefficients $d_n(A)$. However, precisely at such values, $r_n$ deviates sensibly from a linear dependence on $1/n$, and the result from the root test are not reliable.\footnote{Note that at small values of $A$ the sequence $r_n$ splits in two subsequences $r_{2j}$ and $r_{2j+1}$, which differ significantly from each other at small $j$. However we can easily deal with such behavior by fitting appropriate subsequences of $r_n$. On the contrary, in correspondence of the real part of the zeros of a coefficient $d_m(A)$, the corresponding element $r_m$ will deviate from its neighbors, giving rise to an erratic behavior of the sequence $r_n(A)$ which is not easy to take care of.}
In any case we expect that in general the radius of convergence of the series will not extend beyond the fixed singularity at $\Rt_+$, with possibly the exception of accumulation point of the zeros of $d_n(A)$.

\begin{figure}[ht]
\centering \includegraphics[width=12cm]{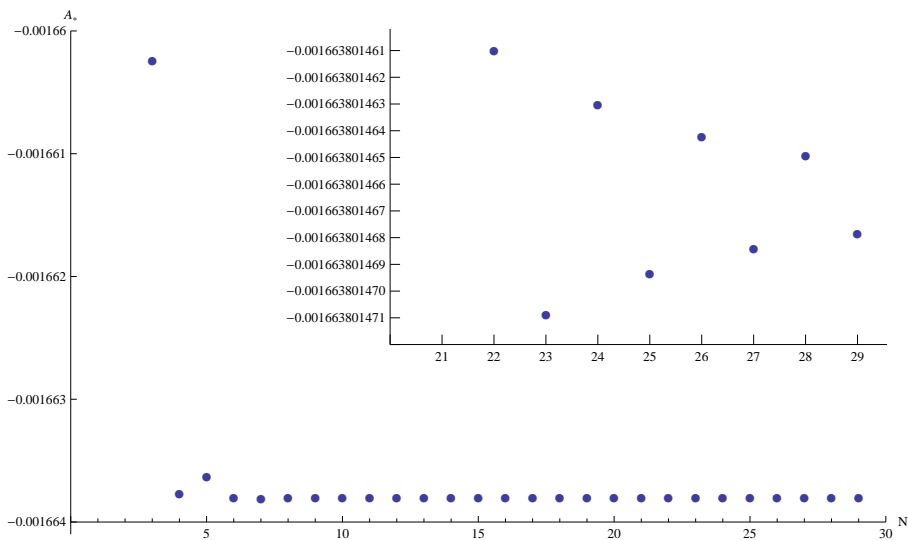}
\caption{
\small{A plot of the most rapidly converging solution to $d_{N+1}=0$, as a function of $N\in\{1,...,29\}$.}
\label{fig:An}
}
\end{figure}

\begin{figure}
\centering
\begin{tabular}{ccc}
\includegraphics[scale=0.52]{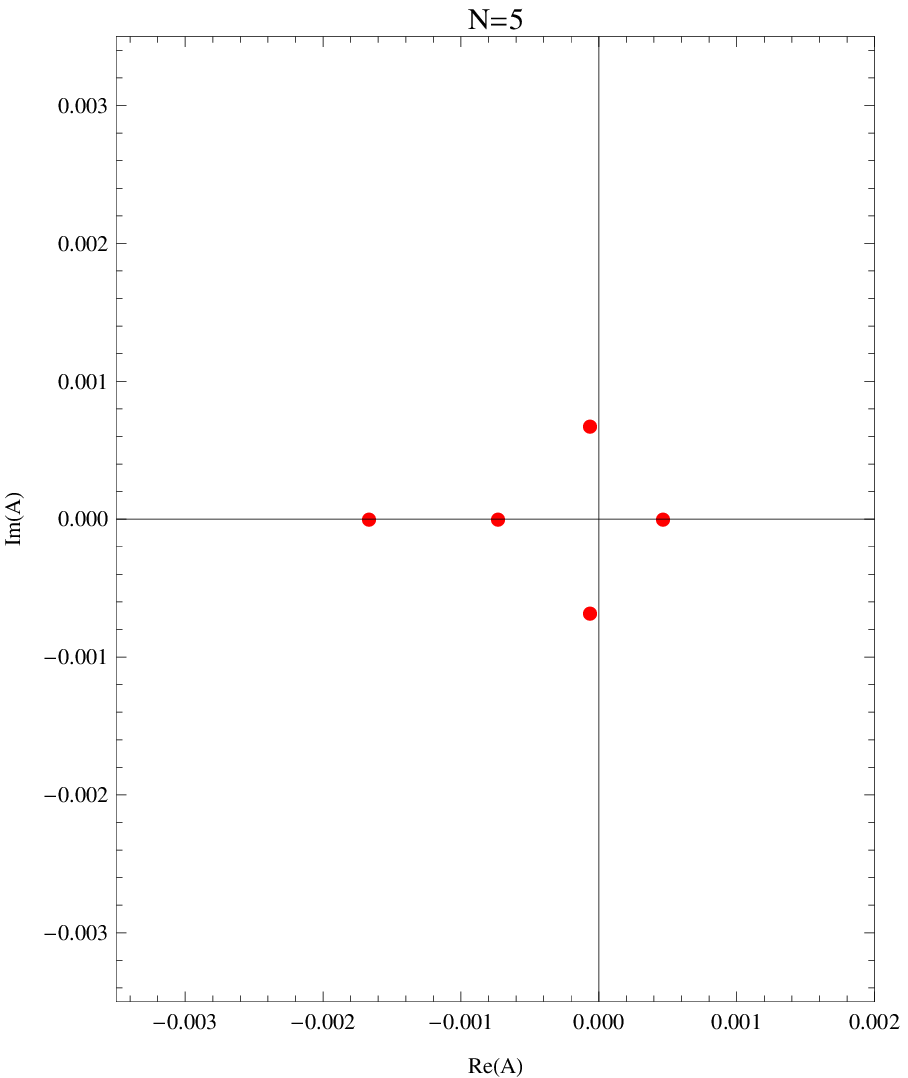} & \includegraphics[scale= 0.52]{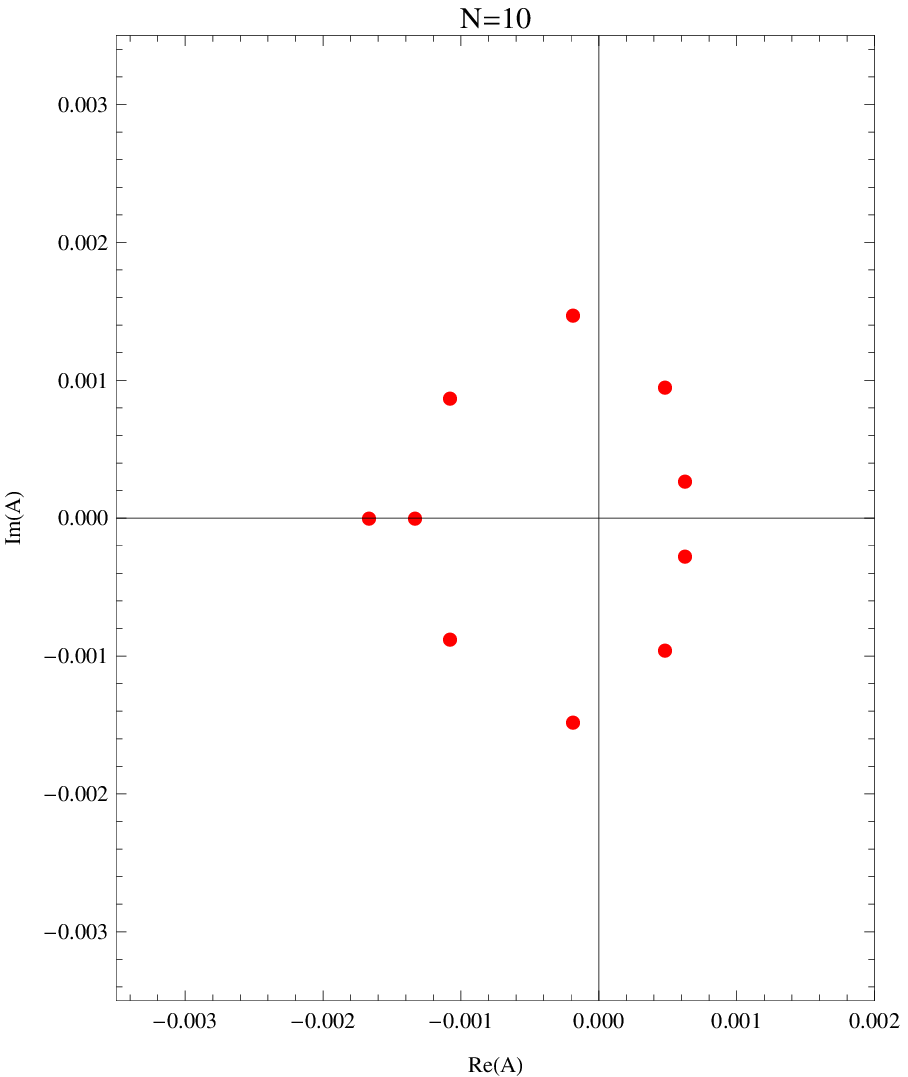} & \includegraphics[scale= 0.52]{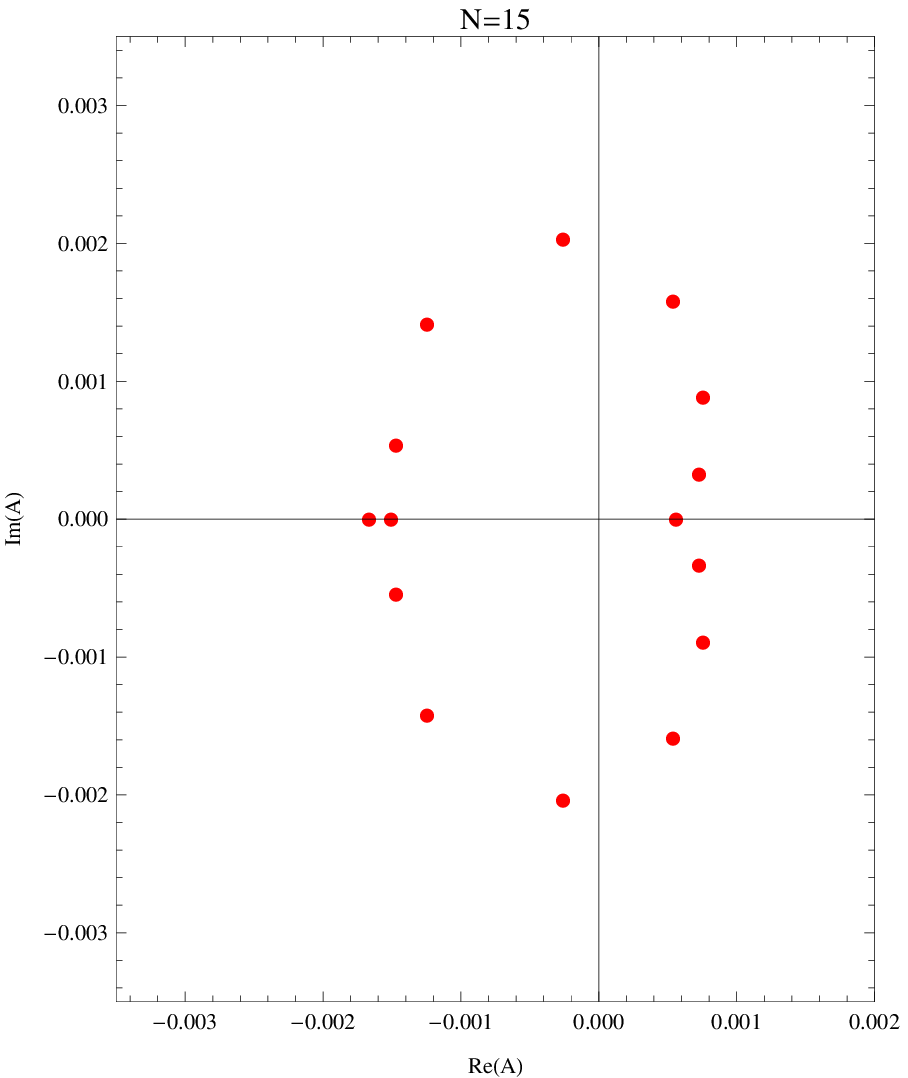} \\
\hline
\includegraphics[scale= 0.52]{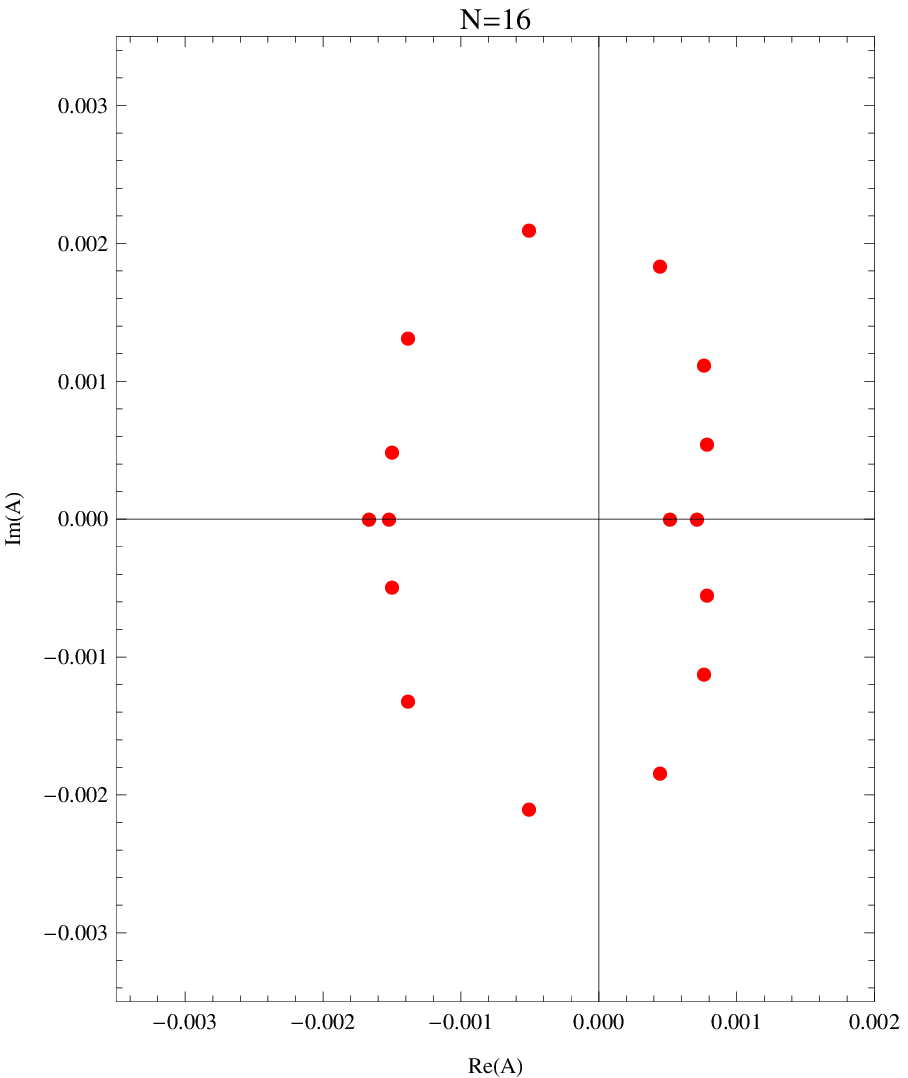} & \includegraphics[scale= 0.52]{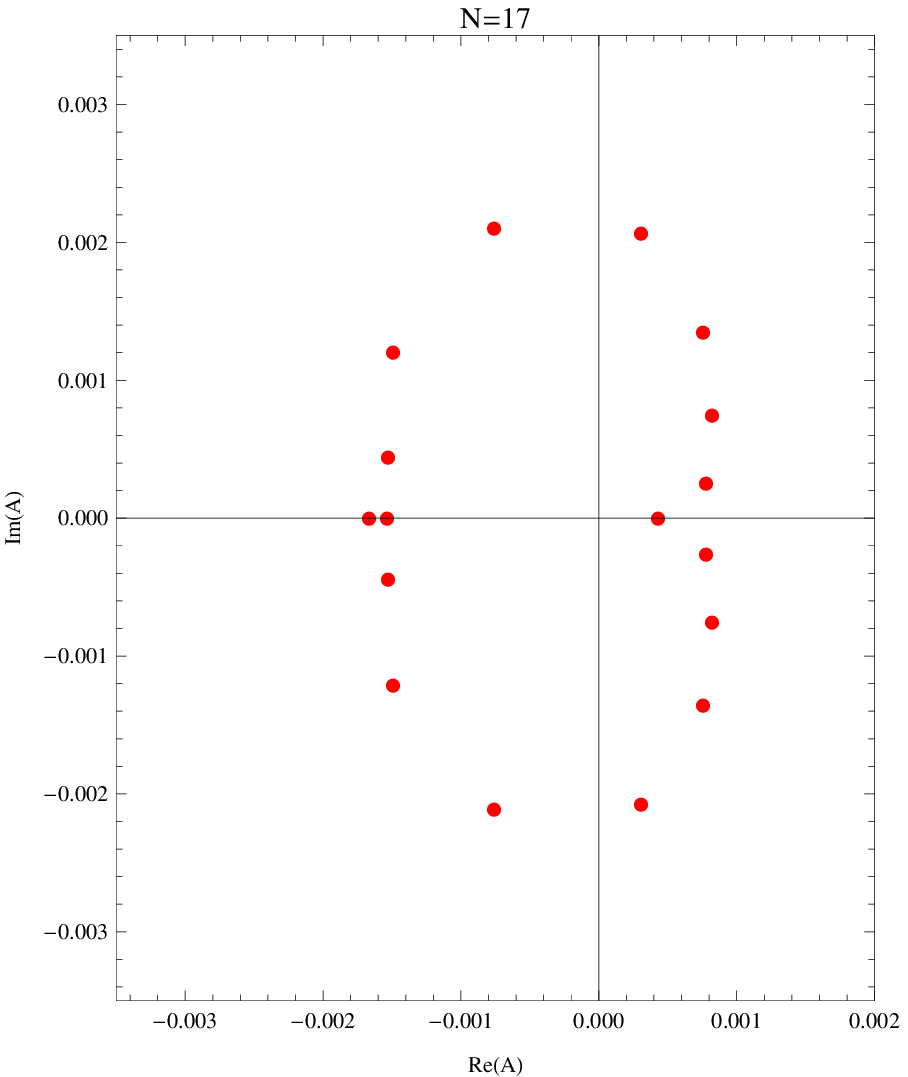} & \includegraphics[scale= 0.52]{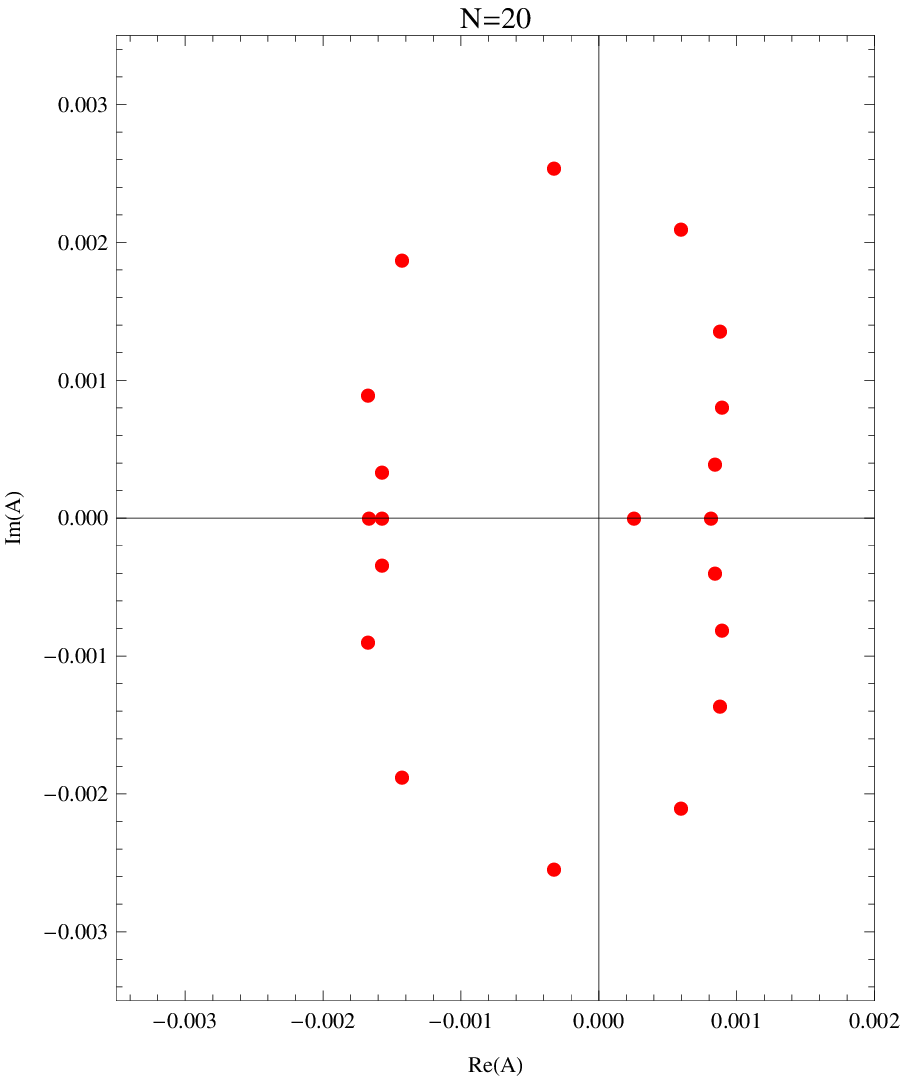} \\
\hline
\includegraphics[scale= 0.52]{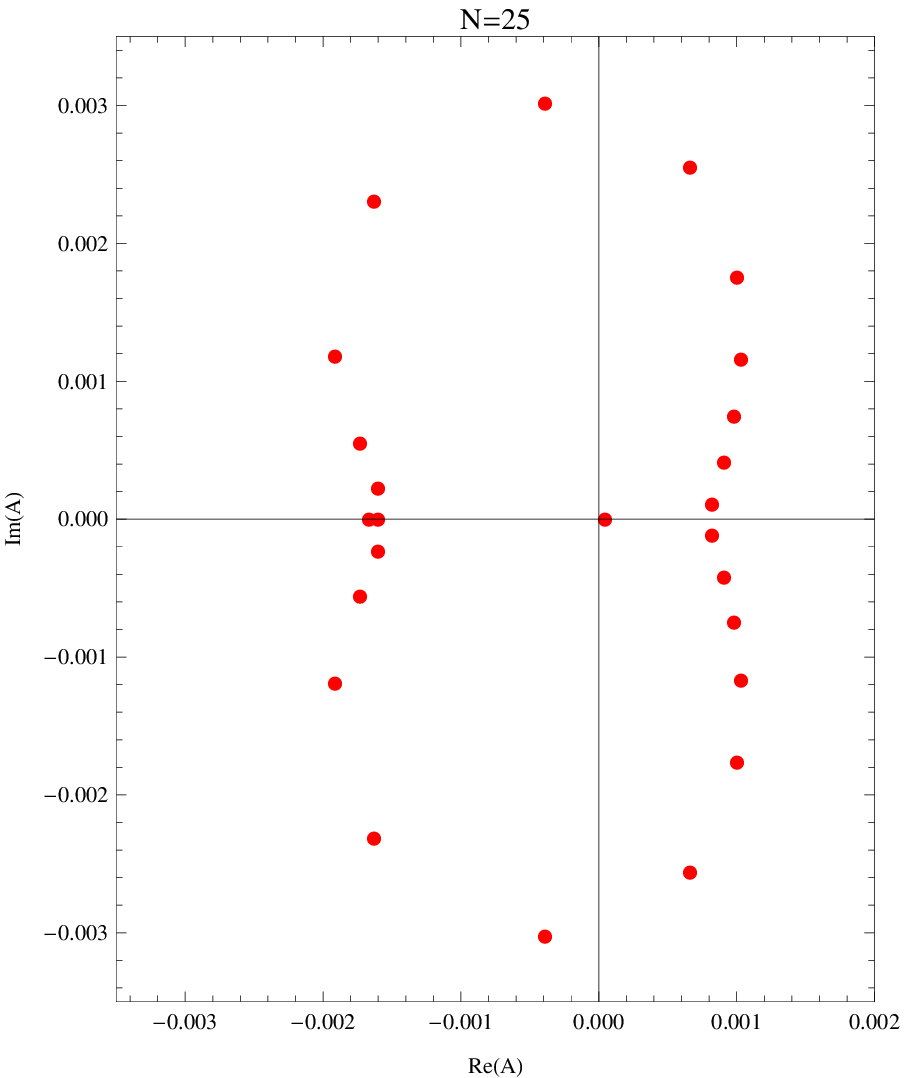} & \includegraphics[scale= 0.52]{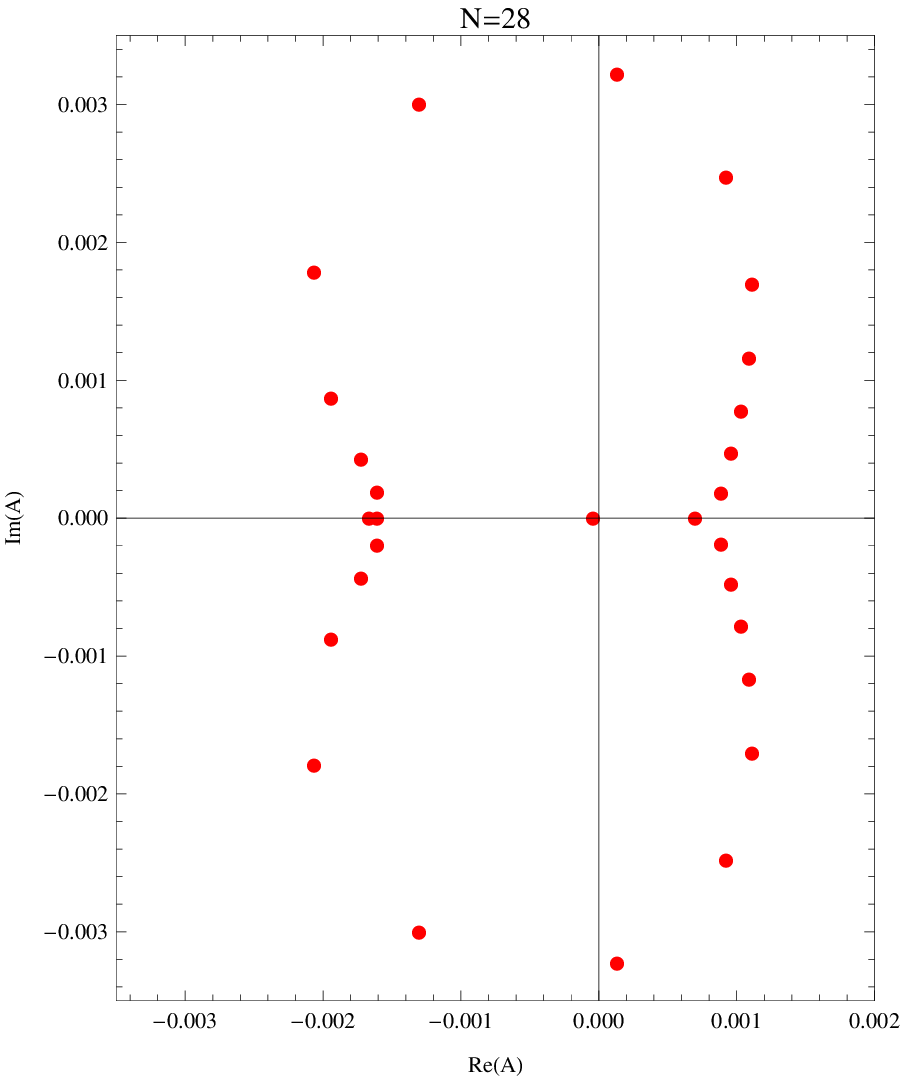} & \includegraphics[scale= 0.52]{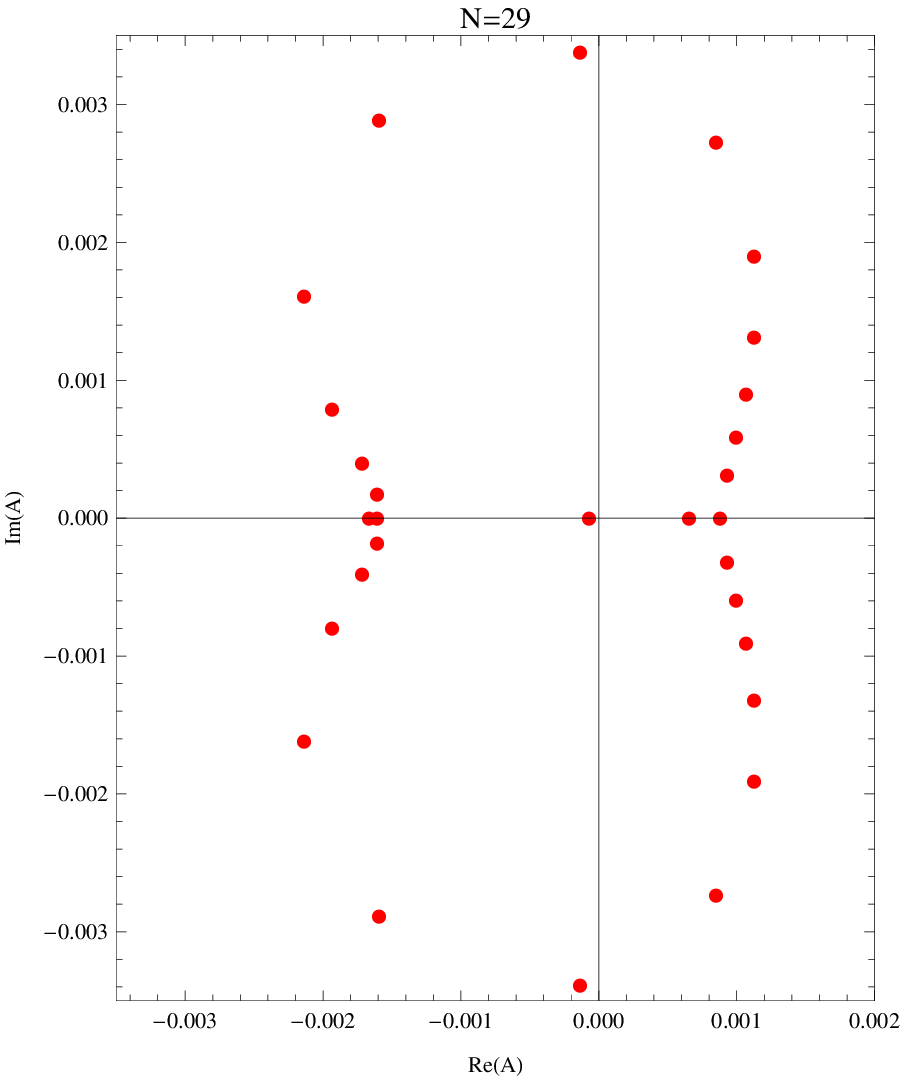} \\
\end{tabular}
\caption{A table of plots of the structure of the fixed points for the large-$\Rt$ truncations. The plots are ordered increasingly from top left to bottom right.
For $N<16$, a positive real fixed point appears at odd values of the truncation order, and it is converted into couples of complex fixed points at even values of $N$. 
At $N=16$ the real fixed point splits into a couple of real fixed point, one of which drifts towards the negative axis, and the other follows the previous patter, with odd and even orders inverted, until it undergoes a new splitting at $N=29$.}
\label{fig:fixpmot}
\end{figure}

An approach equivalent to the standard truncations can be applied also to the asymptotic series \eqref{asymp-sol}. That is, a truncated fixed-point solution of order $N$ is obtained by solving the differential equation up to $d_{N+1}$ and subsequently imposing $d_{N+1}=0$. Of course we find in this way exactly $N$ complex solutions for $A$. It turns out that there is a sequence $A_1(N)$ of zeros which converges very quickly to $A^*=-0.001663801$, as shown in Fig.~\ref{fig:An}, 
while the others move in the complex plane in a characteristic pattern, Fig.~\ref{fig:fixpmot}.

Around a fixed point, we can evaluate the stability exponents $\th_i$
of the beta functions for the coupling of the asymptotic expansion.\footnote{Defining $d_0=A$, and $\p_t d_i=\b_i(\{d_n\})$, the stability matrix is defined as
$ B_{ij}=\partial_j \beta_{i|{A^*}}$. The stability exponents $\{\th_i\}$ are defined as minus the eigenvalues of $B_{ij}$.}
We have computed them for the fixed points of Fig.~\ref{fig:An} up to order $N=8$, and we report them in Table~\ref{critexp}.
It turns out that the number of relevant directions (corresponding to Re$(\th_i)>0$) seems to grow with $N$.

We also computed the stability exponents for the other fixed points at $A<0$, which form another sequence $A_2(N)$ slowly drifting towards the first, and we report
them in Table~\ref{critexp2}. 
Also in this case the number of relevant directions seems to grow with $N$, with only one more irrelevant direction compared to $A_1(N)$, probably corresponding to the trajectory connecting the two fixed points.

\begin{table}
\scriptsize
\begin{center}
\begin{tabular}{|c|c||c|c|c|c|c|c|c|c|c|}\hline
$N$ &  $10^3 A_1$  &  $\th_0$    &  $\th_1$     &   $\th_2$     &     $\th_3$    &       $\th_4$        &     $\th_5$          &    $\th_6$           &     $\th_7$          &       $\th_8$      \\ \hline
1 & -1.319 &-4.26  & 0.65 &       &             &             &             &             &              &\\
2 & -1.631 & 29.2 & 9.19 & -1.64 &             &             &             &             &              &\\
3 & -1.660 & 29.6 & 7.47 & -1.80 & 20.39       &             &             &             &              &\\
4 & -1.663 & 29.4 & 7.25 & -1.82 & 17.3-4.38i & 17.3+4.38i &             &             &              &\\
5 & -1.663 & 29.7 & 7.27 & -1.81 & 14.6-4.24i & 14.6+4.24i & 20.9       &             &              &\\
6 & -1.663 & 29.5 & 7.26 & -1.82 & 14.0-3.19i & 14.0+3.19i & 19.4-4.20i & 19.4+4.20i &              &\\
7 & -1.663 & 29.8 & 7.26 & -1.82 & 14.4-2.50i & 14.4+2.50i & 16.7-4.72i & 16.7+4.72i &  22.5       &\\
8 & -1.663 & 29.5 & 7.26 & -1.82 & 14.8-3.55i & 14.8+3.55i & 15.8-2.32i & 15.8+2.32i &  21.6+3.83i & 21.6-3.83i \\ \hline
\end{tabular}\end{center}
\caption{The critical exponents for the sequence $A_1(N)$ of Fig.~\ref{fig:An}, at the orders $N=1,\cdots,8$.}
\label{critexp}
\end{table}

\begin{table}
\scriptsize
\begin{center}
\begin{tabular}{|c|c||c|c|c|c|c|c|c|c|c|}\hline 
$N$  &  $10^3 A_2$ &  $\th_0$    &  $\th_1$     &   $\th_2$     &     $\th_3$    &       $\th_4$        &     $\th_5$          &    $\th_6$           &     $\th_7$          &       $\th_8$          \\\hline
3   & -0.225 & -19.1  &  9.78 - 5.68i  &  9.78  + 5.68i  &               &                &                  &                  &                  &               \\
4  &  -0.522 & -27.3 &  8.08 - 5.60i  &  8.08  + 5.60i  & -0.51         & 13.0          &                  &                  &                  &                \\
5  &  -0.727 & -29.1 &  7.68 - 5.96i  &  7.68  + 5.96i  & -0.51         & 13.1          & 14.9            &                  &                  &                \\
6  &  -0.910 & -34.8 &  8.55 - 6.05i  &  8.55  + 6.05i  & -0.56         & 13.4          & 15.0 + 1.75i    & 15.0 - 1.75i    &                  &                 \\
7  &  -1.05 & -47.8 &  9.28 - 6.94i  &  9.28  + 6.94i  & -0.61         & 14.2          & 13.8 + 1.41i    & 13.8 - 1.41i    & 17.6            &      \\
8  &  -1.17 & -78.1 &  9.43 - 7.66i  &  9.43  + 7.66i  & -0.67         & 13.5          & 17.5 + 1.09i    & 17.5 - 1.09i    & 14.5            & 15.3            \\\hline
\end{tabular}\end{center}
\caption{The critical exponents for the sequence $A_2(N)$.}
\label{critexp2}
\end{table}

\begin{table}
\scriptsize
\begin{center}
\begin{tabular}{|c|c||c|c|c|c|c|c|c|c|}\hline  
$N$  &  $10^4 A_3$ &     $\th_0$    &  $\th_1$     &   $\th_2$     &     $\th_3$    &       $\th_4$        &     $\th_5$          &    $\th_6$           &     $\th_7$           \\\hline
3 & 2.54 & -0.55 & 6.60           & 4.03 - 10.8i    & 4.03 + 10.8i  &                &                  &                  &  \\
5 & 4.65 & -0.52 & 9.73           & 2.12 - 18.4i    & 2.12 + 18.4i  & 8.55 - 2.71i   & 8.55 + 2.71i     &                  & \\
7 & 5.72 & -0.54 & 8.74        & 9.54 - 4.15i   & 9.54 + 4.15i   & 12.7    & 13.5          & -4.30 - 23.7i   & -4.30 + 23.7i \\\hline
\end{tabular}\end{center}
\caption{The critical exponents for the sequence of real fixed point at $A>0$, appearing at odd values of $N$. 
}
\label{critexp3}
\end{table}

It should be mentioned that truncations with single $R^{-n}$ terms added to the Einstein-Hilbert actions were considered in \cite{Machado:2007ea}, where it was observed that such new terms correspond to new relevant directions. Our results are hence compatible with that of \cite{Machado:2007ea}.

\begin{figure}
\centering
\includegraphics[scale=0.65]{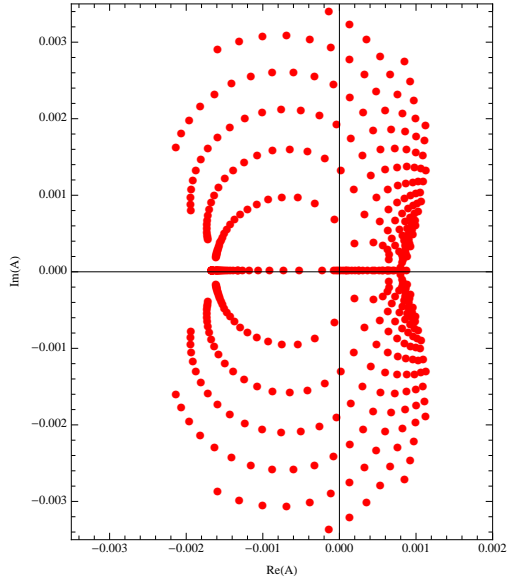}
\caption{A simultaneous plot of all the fixed points for the large-$\Rt$ truncations, at various $N$.}
\label{fig:Atot}
\end{figure}

Finally, it should also be noted that negative values of $A$ correspond to an unbounded fixed-point action (see Sec.~\ref{Sec:FP-action}), and might be not very appealing for this reason.
Potentially interesting candidates for a fixed point could maybe be found also at positive $A$, slightly relaxing the type of convergence one is looking for. 
One way to visualize other interesting points is to show all the solutions for different values of $N$ in the same plot, as in  Fig.~\ref{fig:Atot}.
In this way we can see that there are some accumulation points for the zeros of $d_{N+1}(A)$, and these are potential candidates for global solutions of the differential equation.
Of course if a sequence of zeros converges to a point $A^*$, like in Fig.~\ref{fig:An}, this will be an accumulation point.
On the other hand, accumulation points appear also whenever a fixed point which appears at odd orders, splits into a couple of nearby complex ones at even orders (or vice versa).
One such sequence of points is present in our solutions, and in Table~\ref{critexp3} we list the relative stability exponents. However, due to the fact that only odd orders appear, and that higher orders are very demanding from a computational point of view, the available data are not enough to establish any convergence of its critical exponents.

\subsection{Numerical integration}
\label{Sec:numint}

One advantage of having a differential equation for the fixed-point effective action is the possibility of going beyond the series expansions discussed so far, and to study its solutions via numerical integration methods.
In principle we would like to employ such methods in order to look as in \cite{Morris:1994ki} for the initial conditions leading to global solutions.\footnote{At least solutions defined on the whole positive axis plus the origin. As the equation \eqref{FP-eq2} was derived assuming a spherical background, we might not trust it at negative $\Rt$.} Unfortunately, the presence of two free initial conditions at the origin, combined with complicated structure of the equation, and the presence of a fixed singularity at $\Rt_\pm$, renders such an analysis much more involved than in the scalar case. We report here on some preliminary results in this direction.

\begin{figure}[ht]
\centering 
\begin{tabular}{cc}
\includegraphics[width=6cm]{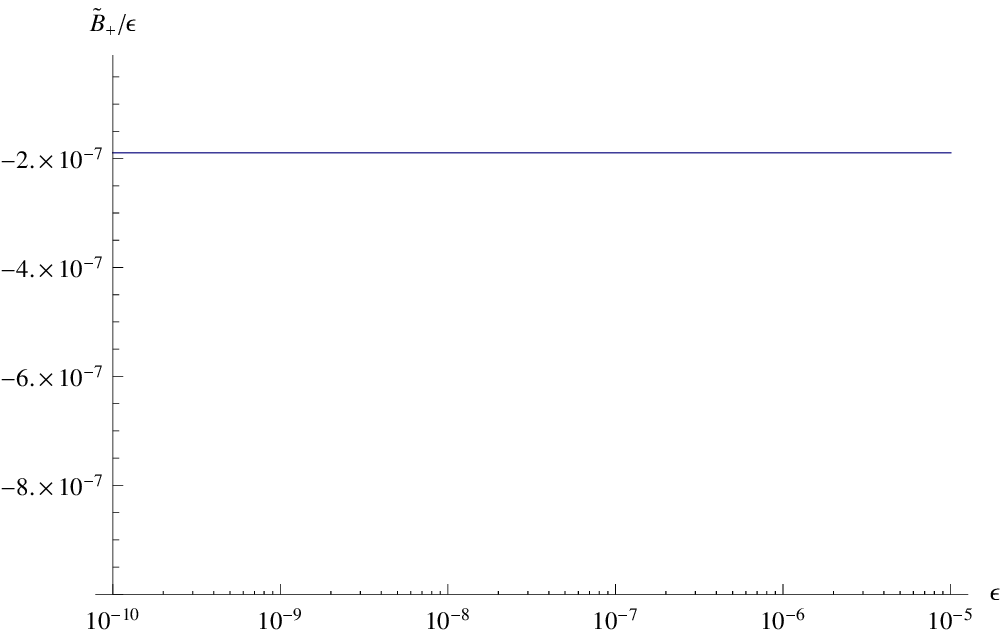} & \hspace{.7cm} \includegraphics[width=6cm]{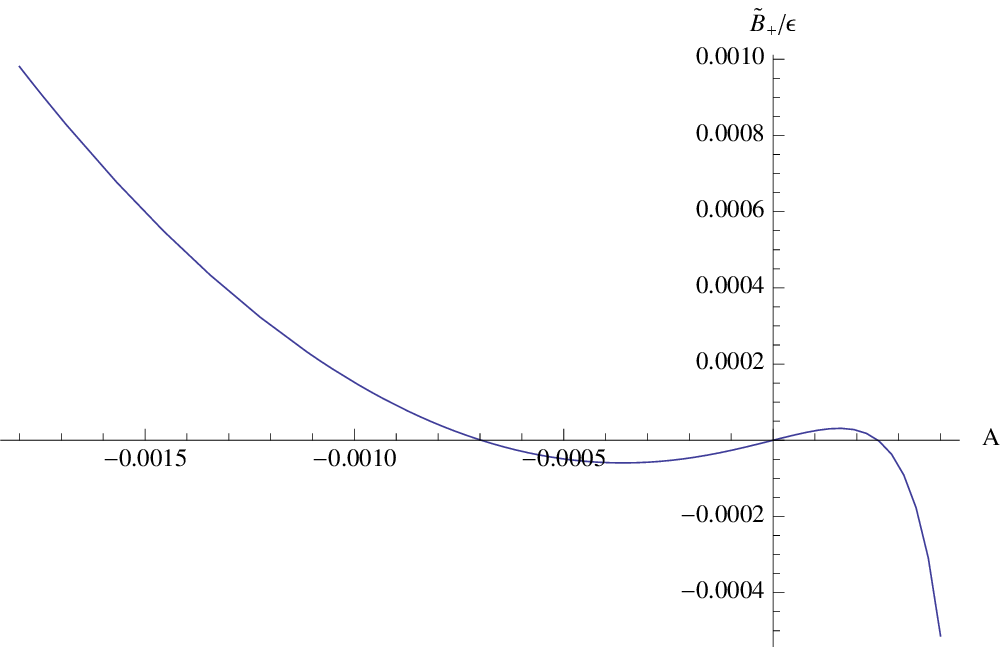} \\
\end{tabular}
\caption{
\small{On the left we show a plot of $\tilde{B}_+(b_0,b_1,b_2)/\eps$ as a function of $\eps$ at $A=.00015$ (similar plots are obtained in the whole range $-0.0035\lesssim A\lesssim0.0005$). As trivial as it might look like, the plot carries an important message: $\tilde{B}_+(b_0,b_1,b_2)\sim\eps$ at these values of $A$.
The plot on the right shows instead $\tilde{B}_+(b_0,b_1,b_2)/\eps$ as a function of $A$: various plots for different values of $\eps$ are shown simultaneously, and the superposition is perfect.}
\label{fig:Btplus_of_eps}
}
\end{figure}

The most efficient approach to the numerical investigations is to start integration at large $\Rt$, where the series expansion presents only one free parameter.\footnote{The starting value of $\Rt$ has to be chosen within the convergence radius of the series in \eqref{asymp-sol}, and also for this reason we have detailed the convergence analysis in the previous section.} Imposing initial conditions matching \eqref{asymp-sol}, and integrating backward towards small $\Rt$, we can monitor the appearance of singularities as we vary the parameter $A$.
Naively we would like to vary $A$ in order to find those values for which we can integrate all the way to $\Rt=0$ without running into a singularity of the type \eqref{mov-sing}.
However, because of the singularity at $\Rt_+$ we expect that to be impossible, as indeed confirmed by explicit numerical integration.
\begin{figure}[ht]
\centering \includegraphics[width=9cm]{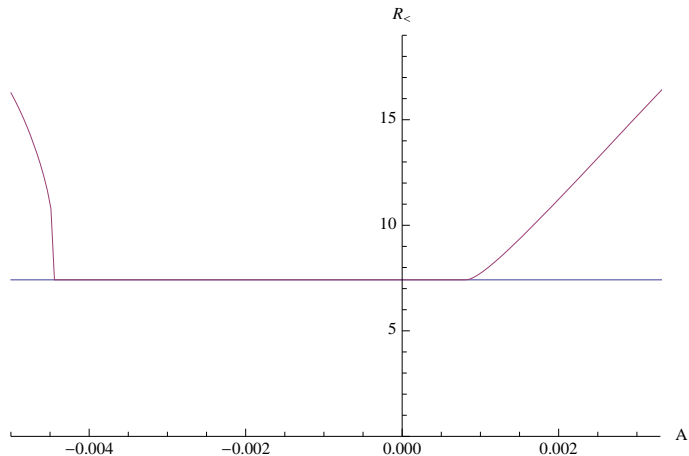}
\caption{\label{fig:numsingA1}
\small{A plot of the final integration point $\Rt_<$ reached numerically (starting at $\Rt=60$), i.e. of the location of the movable singularity, as a function of $A$.
The horizontal line is $\Rt=\Rt_+$.}
}
\end{figure}
The plot Fig.~\ref{fig:numsingA1} shows that the numerical integration breaks down once the fixed singularity at $\Rt_+$ is reached.
As we already know from Sec.~\ref{Sec:Rt+}, in order to have a regular solution at $\Rt_+$ the condition  $\tilde{B}_+(b_0,b_1,b_2)=0$ must be satisfied, which hence we would like to impose as a condition on $A$.
However the numerical integration can only be carried out until an $\eps$ away from $\Rt_+$. We have thus evaluated and plotted $\tilde{B}_+(b_0,b_1,b_2)$ at $\Rt_+ +\eps$ as a function of $\eps$ for various values of $A$, in order to extrapolate the result to $\eps=0$. It turns out that $\tilde{B}_+(b_0,b_1,b_2)$ scales to zero with $\eps$ in the range $-0.0035\lesssim A\lesssim0.0005$, while it diverges outside of that (if it reaches $\Rt_+$ at all).
As an example of scaling we report a plot of $\tilde{B}_+(b_0,b_1,b_2)/\eps$ as a function of $\eps$ and $A$ in Fig.~\ref{fig:Btplus_of_eps}.
As a result, the regularity condition at $\Rt_+$ does not fix $A$ to a discrete set but just to a continuous interval.

Within such interval, numerical integration could be continued beyond $\Rt_+$ by matching the numerical solution at  $\Rt_++\eps$ with the series solution 
\eqref{serieR+}, and use the latter to impose initial conditions at $\Rt_+-\eps$ for a new numerical integration towards $\Rt=0$. 
We would have to look for the range of $A$ which allows integration down to $\Rt=\eps$ and here impose the analyticity constraint \eqref{B0}, again in the limit $\eps\to 0$.
A number of technical difficulties appear along this path, in particular due to a very slow convergence of the series \eqref{serieR+}, and a detailed implementation of this program is left for future work.

However we have achieved an important result: the mismatch between number of conditions and number of parameters (two analyticity conditions for one free parameter) does not seem to provide an obstruction for the existence of global solutions.
The condition at $\Rt=\Rt_+$ leaves us still with a continuous degree of freedom, hence anything could happen when imposing the condition at $\Rt=0$: we could find a discrete set (eventually empty) of solutions, as well as another continuous interval.

\subsection{Fixed point action}
\label{Sec:FP-action}

One very general conclusion can be easily derived from the large $\Rt$ series expansion studied in Sec.~\ref{Sec:infty}: 
if \eqref{FP-eq2} admits a global fixed-point solution $\ft^*(\Rt)$, then necessarily
\be \label{FP-EA}
\G^*=\G^*_{k=0}=A^* \int d^4 x\sqrt{g}\, R^2 \, ,
\ee
for some finite $A^*$.
That is, the fixed-point effective action corresponds to an interacting $R^2$ theory.

Here the effective action is the standard one, obtained from the average effective action in the limit $k\to 0$, i.e. it is the effective action obtained by a path integral with no infrared cutoff.
Obtaining the effective action is generally a difficult task, which would require integrating the FRGE down to $k=0$, starting from some initial condition at some $k_0>0$ (e.g. at the UV cutoff, ideally with an initial condition on a trajectory emanating from the UV fixed point, to ensure independence from the UV cutoff).
However, at the fixed point such integration is trivial, as the $k$-dependence of $\G^*_k$ is contained in a trivial scaling. We can easily prove \eqref{FP-EA} by noticing that
\be \label{FP-EAk}
\G^*_k= k^4 \int d^4 x\sqrt{g}\, \ft^*(R/k^2) \, ,
\ee
and hence the limit $k\to 0$ corresponds to the limit $\Rt\to\infty$ in $\ft(\Rt)$.
Inserting \eqref{asymp-sol}, with $A$ at the fixed-point value $A^*$, into \eqref{FP-EAk}, and taking the limit $k\to 0$, we find \eqref{FP-EA}.
Note that we had already met an $R^2$ action in the case of the Gaussian fixed point. However, in that case $A^*\to\infty$, requiring a rescaling of the fluctuation field, and a reduction to a free theory.

The result \eqref{FP-EA} could be expected on dimensional grounds, from the simple fact that the fixed-point action should be scale invariant.
However, naive scaling of a term of the Lagrangian is only valid if it has no anomalous dimension. 
For example,  in $d=3$ the scalar potential at the Wilson-Fisher fixed point is $V(\phi)=A^* \phi^{\f{6}{1+\eta}}$, and $\eta=0$ in the LPA, but $\eta\sim0.03$ at higher orders of the derivative expansion \cite{Litim:2010tt}.
What we found here is one more analogy between the $f(R)$ approximation and the LPA: within the $f(R)$ approximation the anomaouls dimension of $R$ is zero and hence the fixed-point effective action is an $R^2$ action.

Furthermore, such result was impossible to see in the polynomial truncations studied in previous works: the results from such truncations \cite{Codello:2007bd,Machado:2007ea,Codello:2008vh,Bonanno:2010bt} indicate that every coupling in the expansion is non-zero at the NGFP, and no hint on how the series would sum up was available.

\section{Conclusions}
\label{Sec:concl}

In this article we have derived and studied a differential equation for the renormalization group fixed points of gravity in the $f(R)$ approximation. We have argued that such approximation plays a role analogous to the one played by the local potential approximation in scalar field theories, and taking seriously such perspective we have examined various properties of the solutions to our differential equation. In particular we have studied regularity conditions and large-$\Rt$ expansion. Within the latter we have checked the convergence of fixed-point solutions obtained by truncations of the series expansion.

The main results of our analysis are the following:
\begin{itemize}
\item the fixed-point structure of large-$\Rt$ truncations is much clearer than the one from small $\Rt$ truncations, and in particular we found striking convergence of a fixed-point solution (Fig.~\ref{fig:An} and \ref{fig:fixpmot}), which however turns out to have a growing number of relevant directions;
\item the analyticity condition at $\Rt=\Rt_+$ is satisfied on a continuous interval of the parameter $A$, hence the presence of two fixed singularities in our equation is not an obstruction to the existence of global solutions;
\item if global solutions exist, they necessarily correspond to the effective action of an $R^2$ theory.
\end{itemize}

The main motivation for our investigations was the asymptotic safety scenario conjectured by Weinberg, and our hope is that the approach we have presented here will help understanding further the status of such conjecture.
It should be pointed out however that our methods and results are quite generic and could also be useful for the investigation of other aspects of gravity, such as infrared modifications, stability of the de Sitter solution, and nucleation of black holes (see \cite{Cognola:2005de} for a one-loop analysis of such questions within an $f(R)$ setting).

What we presented here was just a first step in the direction of exploiting the $f(R)$ approximation in the spirit of the local potential approximation, and many more developments of this approach are possible and needed. For example, on a technical side, the use of a different cutoff would be desirable, possibly avoiding the staircase nature of the traces.
Pushing the Table~\ref{critexp3} to higher orders would be important for testing the existence of a fixed point with $A^*>0$ and with a finite number of relevant directions.
And of course, a more comprehensive exploration of the numerical solutions is being performed, and it will be the subject of a future publication.

Finally, the result of Sec.~\ref{Sec:FP-action} suggests that in a more generic approximation (i.e. not relying on a maximally symmetric background) the fixed-point action would probably contain a Weyl-squared term. As a consequence, we feel that the issue of unitarity of the resulting theory will need to be discussed further, for example sharpening the argument presented in \cite{BMS1}.

\begin{center}
\vspace{.4cm}
{\bf Aknowledgements}\\  
\end{center}

Research at Perimeter Institute is supported by the Government of Canada
through Industry Canada and by the Province of Ontario
through the Ministry of Research \& Innovation.


\providecommand{\href}[2]{#2}\begingroup\raggedright\endgroup

\newpage

\section{Erratum}

Equation \eqref{W0} in the original paper contains a mistake, and it should be corrected to
\be
\begin{split}
W_0^{\hb}(z,\Rt) = & \frac{3 (1-z^2) \left(3 \p_t\ft''(\Rt)-6 \Rt   \ft^{(3)}(\Rt)\right)}{18 \ft''(\Rt)-2 (\Rt-3)   \ft'(\Rt)+4 \ft(\Rt)}
 \\   &   +\frac{3 (1- z )  \left(\p_t\ft'(\Rt)-2 \Rt \ft''(\Rt)+2   \ft'(\Rt)\right)+6\left(\ft'(\Rt)+6\ft''(\Rt)\right)}{18 \ft''(\Rt)-2 (\Rt-3)   \ft'(\Rt)+4 \ft(\Rt)} \, .
\end{split}
\ee
The error propagates to Sections \ref{Sec:FPeq} and \ref{Sec:analysis}, affecting part of the numerical results, but not the qualitative aspects.
In particular, equation \eqref{Th} should be
\be
\begin{split}
\cT^{\hb}_0 = & \frac{ 1}{
   2 \Rt^2 \left(-9 \ft''(\Rt)+(\Rt-3) \ft'(\Rt)-2
   \ft(\Rt)\right)
   } \times \\
&\quad \Big\{ \left(\Rt^4-54 \Rt^2-54\right) \left( \p_t\ft''(\Rt) -2\Rt  \ft^{(3)}(\Rt) \right) \\
&\qquad  -\left(\Rt^3+18 \Rt^2+12\right) \left(  \p_t\ft'(\Rt)     -2\Rt \ft''(\Rt) + 2    \ft'(\Rt) \right) \\
&\qquad -36 \left(\Rt^2+2\right) \left(\ft'(\Rt)+6\ft''(\Rt)\right) \Big\} \, .
\end{split}
\ee
Equations \eqref{eq-numerator} and \eqref{B+} should be corrected accordingly, as well as the numerical results in Sections \ref{Sec:infty} and \ref{Sec:numint}. All the qualitative aspects of the results (e.g. Section \ref{Sec:FP-action}), as well as the general assessment and conclusions, are unaffected by such corrections.

\end{document}